\documentclass{aa}  
\usepackage{graphicx}
\usepackage{eqnarray,amsmath}
\usepackage{multicol}
\usepackage[hypcap]{caption}
\usepackage{txfonts}
\usepackage{hyperref}
\usepackage{xcolor}
\graphicspath{{./figures/}} % Location of the graphics files
\begin{document} 

\title{New self-consistent wind parameters to fit optical spectra of O-type stars observed with the HERMES spectrograph}
\titlerunning{Wind parameters to fit optical spectra observed with the HERMES spectrograph}

%   \subtitle{I. Overviewing the $\kappa$-mechanism}

\author{A. C. Gormaz-Matamala\inst{1,2,3}
\and
M. Curé\inst{2,3}
\and
A. Lobel\inst{4}
\and
J. A. Panei\inst{5,6}
\and
J. Cuadra\inst{1}
\and
I. Araya\inst{7}
\and
C. Arcos\inst{2,3}
\and
F. Figueroa-Tapia\inst{2}}
%\fnmsep\thanks{Just to show the usage of the elements in the author field}

\institute{Departamento de Ciencias, Facultad de Artes Liberales, Universidad Adolfo Ib\'a\~nez, Av. Padre Hurtado 750, Vi\~na del Mar, Chile\\
\email{alex.gormaz@uv.cl}
\and
Instituto de Física y Astronomía, Universidad de Valparaíso. Av. Gran Breta\~na 1111, Casilla 5030, Valpara\'iso, Chile.
\and
Centro de Astrofísica, Universidad de Valparaíso. Av. Gran Breta\~na 1111, Casilla 5030, Valpara\'iso, Chile.
\and
Royal Observatory of Belgium, Ringlaan 3, 1180 Brussels, Belgium
\and
Departamento de Espectroscop\'{\i}a, Facultad de Ciencias Astron\'omicas y Geof\'{\i}sicas, Universidad Nacional de La Plata (UNLP), Paseo del Bosque S/N (B1900FWA), La Plata, Argentina.
\and
Instituto de Astrof\'{\i}sica de La Plata, CCT La Plata, CONICET-UNLP, Paseo del Bosque S/N (B1900FWA), La Plata, Argentina.
\and
Centro de Investigación DAiTA Lab, Facultad de Estudios Interdisciplinarios, Universidad Mayor, Alonso de Córdova 5495, Santiago, Chile.}

\date{Received xxx; accepted XXX}

% \abstract{}{}{}{}{} 
% 5 {} token are mandatory
 
\abstract % 5 {} token are mandatory
% context heading (optional)
{}
% aims heading (mandatory)
{We performed a spectral fitting for a set of O-type stars based on self-consistent wind solutions, which provide mass-loss rate and velocity profiles directly derived from the initial stellar parameters.
The great advantage of this self-consistent spectral fitting is therefore the reduction of the number of free parameters to be tuned.}
% methods heading (mandatory)
{Self-consistent values for the line-force parameters $(k,\alpha,\delta)_\text{sc}$ and subsequently for the mass-loss rate, $\dot M_\text{sc}$, and terminal velocity, $\varv_{\infty,\text{sc}}$, are provided by the m-CAK prescription introduced in Paper I, which is updated in this work with improvements such as a temperature structure $T(r)$ for the wind that are self-consistently evaluated from the line-acceleration.
Synthetic spectra were calculated using the radiative transfer code FASTWIND, replacing the classical $\beta$-law for our new calculated velocity profiles $\varv(r)$ and therefore making clumping the only free parameter for the stellar wind.}
% results heading (mandatory)
{We found that self-consistent m-CAK solutions provide values for theoretical mass-loss rates of the order of the most recent predictions of other studies.
From here, we generate synthetic spectra with self-consistent hydrodynamics to fit and obtain a new set of stellar and wind parameters for our sample of O-type stars (HD 192639, 9Sge, HD 57682, HD 218915, HD 195592, and HD 210809), whose spectra were taken with the high-resolution echelle spectrograph \textsc{Hermes} ($R=85\,000$).
We find a satisfactory global fit for our observations, with a good accuracy for photospheric He I and He II lines and a quite acceptable fit for H lines.
Although this self-consistent spectral analysis is currently constrained in the optical wavelength range alone, this is an important step towards the determination of stellar and wind parameters without using a $\beta$-law.
Based on the variance of the line-force parameters, we establish that our method is valid for O-type stars with $T_\text{eff}\ge30$ kK and $\log g\ge3.2$.
Given these results, we expect that the values introduced here are helpful for future studies of the stars constituting this sample, together with the prospect that the m-CAK self-consistent prescription may be extended to numerous studies of massive stars in the future.}
% conclusions heading (optional), leave it empty if necessary 
{}

\maketitle

%_____INTRODUCTION_______________________________________________________________________________
\section{Introduction}
	Massive stars strongly influence the interstellar medium by means of their winds, which carry momentum, energy, and chemical elements \citep[see reviews from][]{kudritzki00,puls08}.
	The study of their stellar winds is then an important topic in stellar astrophysics because it connects the physics of the mechanism generating the outflows of massive stars with the observed phenomena at a Galactic level, such as stellar evolution –and subsequently, stellar populations– or gas dynamics at the Galactic centre \citep{cuadra08}, and even at a extragalactic level, such as the so-called wind-momentum-luminosity relation \citep{puls96,kudritzki99}.

	Through the pioneering work of \citet{lucy70}, we know that the winds of hot stars are accelerated by the radiation field emanating from the photosphere of the star.
	Momentum is delivered to the wind primarily by scattering (absorption or re-emission) of photons in bound-bound (i.e. line) transitions.
	A quantitative description of line-driven winds was first performed by \citet[][called the CAK theory]{cak} and was later improved by \citet{abbott82}, \citet{ppk}, and \citet[][called modified or m-CAK theory]{friend86}.
	The approximations and assumptions underlying m-CAK and their implications for the wind dynamics have been extensively discussed \citep{schaerer94,puls08,kk10}; the most important approximation is the Sobolev approximation \citep{sobolev60,castor70}, which allows a description of the line-acceleration in terms of local wind quantities and the photospheric flux \citep{cak,lamersandcassinelli}.

	The beauty of m-CAK theory is that it provides key insights into the relation between the line-force parameters $(k,\alpha,\delta)$ and fundamental parameters describing the atomic transitions responsible for driving the flow, such as their $gf$-values and their line-statistics \citep[see][]{puls00}.
	Moreover, these line-force parameters provide a solution for the wind hydrodynamics, which is beyond the so-called $\beta$-law.
	Another approach for studying stellar winds is to use the Sobolev approximation in conjunction with Monte-Carlo simulations.
	One of the most well-known studies is the Monte-Carlo simulations performed by \citet{vink00,vink01}, whose theoretical mass-loss rates have been widely used to calculate evolutionary tracks \citep{georgy12,ekstrom12}.
	Unfortunately, improved observational studies have shown discrepancies between these Vink rates and observations \cite[e.g.,][]{bouret05,bouret12}.

	In parallel, a renewed effort has been made in past years to improve our understanding of line-driven winds beyond the m-CAK.
	\citet{sander17} calculated consistent line-acceleration and hydrodynamics for the particular case of $\zeta$-Puppis using the PoWR code \citep{grafener02,hamann03}, whereas \citet[][hereafter KK17]{kk17} provided new confident theoretical values for wind parameters of O dwarfs, giants, and supergiants by solving the radiative transfer equation and hydrodynamics simultaneously using the Sobolev approximation for transition rates and co-moving frame (CMF) for the calculation of the radiative force.
	We also mention the recent studies of \citet{sundqvist19} and \citet{bjorklund21}, who used their full non-local (NLTE) CMF radiative transfer solutions from the code FASTWIND \citep{santolaya97,puls05,sundqvist18} to calculate the radiative acceleration responsible for wind driving.
	In the same line, we highlight the study by \citet{alex21a}, who also performed a self-consistent calculation of hydrodynamics and radiative acceleration under a full NLTE treatment using CMFGEN \citep{hillier90b,hillier90a,hillier98}.
	However, all these works require a considerable computational effort, which at the moment prevents the creation of a large grid of self-consistent solutions for a wide range of temperatures and masses.

	Based on the m-CAK theory, \citet[][hereafter Paper I]{alex19} developed a prescription to derive self-consistent solutions given different sets of stellar parameters for hot massive stars, following the quasi-NLTE treatment for atomic populations of \citet{mazzali93}, the radiation field calculated from \textsc{Tlusty} \citep{hubeny95}, and the hydrodynamic solutions performed by \textsc{HydWind} \citep{michel04,michel07}. 
	Likewise, \citet{araya17} self-consistent velocity profiles calculated from this m-CAK prescription can be introduced in FASTWIND to calculate synthetic spectra without any $\beta$-law.
	Paper I demonstrated that this prescription provides useful theoretical wind parameters for different sets of temperature, surface gravities, abundances, and even the rotational velocity, whose derived synthetic spectra quickly approach a fair fit.
	Moreover, the line-force parameters have recently been used to cover a broader range of temperatures, although an LTE scenario was assumed and an optical depth was considered that are inconsistent with the hydrodynamics \citep{lattimer21}.
	In this work, we proceed in the usage of the m-CAK prescription performed by Paper I and it to the spectral fitting of a sample of late-type O stars observed by the \textsc{Hermes} spectrograph.
	From this spectral analysis, we obtain new stellar and wind parameters for the stars of the sample (HD 192639, 9Sge, HD 57682, HD 218915, HD 195592, and HD 210809), which are compared with previous studies.
	We also introduce small improvements in the calculation of the line-force parameters, such as a stratified temperature profile and a more robust handling of the errors for our solutions.

	This paper is organised as follows.
	Details of these improvements and a brief description of their implications are given in Section~\ref{mcakprescription}.
	Details of the \textsc{Hermes} spectrograph and the spectra of the sample are provided in Section~\ref{hermesdata}.
	A description of the steps required to obtain the synthetic spectra from the self-consistent solutions is given in Section~\ref{methodology}.
	The results, with the new stellar and wind parameters measured for our sample, are presented in Section~\ref{spectralresults}.
	Finally, Sections~\ref{discussion} and \ref{conclusions} offer the discussion and conclusions of our work, respectively.

\section{Self-consistent solutions under m-CAK prescription}\label{mcakprescription}
	The study performed in Paper I provided an exhaustive analysis of the line-force parameters $k$, $\alpha,$ and $\delta$ from the CAK theory \citep{cak,abbott82}, calculating them self-consistently with the hydrodynamics derived for the wind by the code \textsc{HydWind} \citep{michel04}.
	We call  the velocity $\varv(r)$ and density $\rho(r)$ profiles that describe the wind wind hydrodynamics.
	These two profiles are closely coupled by the equation of momentum,

	\begin{equation}\label{motion1}
		\varv\frac{d\varv}{dr}=-\frac{1}{\rho}\frac{dp}{dr}-\frac{GM_\text{eff}}{r^2}+g_\text{es}\,k\,t^{-\alpha}\left(\frac{N_\text{e}}{W}\right)^\delta\;\;,
	\end{equation}
	where the term $\propto dp/dr$ is the dependence on the pressure gradient, and the term $\propto M_\text{eff}/r^2$ is the dependence on the gravitational force.
	The last term corresponds to the line-acceleration, where the line-force parameters $(k,\alpha,\delta)$ describe the force multiplier

	\begin{equation}\label{forcemultiplier}
		\mathcal M(t)=k\,t^{-\alpha}\left(\frac{N_\text{e}}{W}\right)^\delta=\frac{g_\text{line}}{g_\text{es}}\;,
	\end{equation}
	which is given this name because it multiplies the acceleration due to electron scattering, $g_\text{es}$.
	We note that $\mathcal M(t)$ is presented as a function of $t$, the CAK optical depth

	\begin{equation}\label{t}
		t=\sigma_\text{es} \varv_\text{th}\,\rho(r)\left(\frac{d\varv}{dr}\right)^{-1}\;,
	\end{equation}
	with $\sigma_\text{es}$ being the electron scattering opacity and $\varv_\text{th}$ the mean thermal velocity of the protons of the wind.
	This $t$ term is defined by the CAK theory \citep{cak} and it only depends on the wind structure, differently from the usual optical depth $\tau$, which depends on the structure and composition of the wind.
	The force multiplier also depends on the ionisation density $N_e/W$, which corresponds to the electron density divided by the geometrical dilution factor $W$.

	Hence, $k$, $\alpha,$ and $\delta$ are calculated by fitting $\mathcal M(t)$ over a 2D plane formed by the CAK optical depth $t$ and the ionisation density $N_\text{e}/W,$  where $t$ is constrained in the range in which the Sobolev approximation can be assumed (i.e. the inner limit is the sonic point of the wind, and the outer limit is the infinite).
	As previously noted in Paper I, the optical depth $t$ decreases at infinitum, but never reaches zero.
	The reason is that at larger distances, both density profile and velocity gradient in Eq.~\ref{t} become proportional to $\sim r^{-2}$, so that both factors cancel each other out.
	The meaning and effects of each of the line-force parameters over the line-acceleration and subsequently over the wind parameters are summarised in \citet[][section~2]{alex19}.
	In addition to Eq.~\ref{motion1}, the velocity and density profiles are also connected to each other by means of the mass-conservation equation, where the term for the mass-loss rate, $\dot M$, is introduced,

	\begin{equation}\label{massconservation}
		\rho(r)=\frac{\dot M}{4\pi r^2\varv(r)}\;.
	\end{equation}

\begin{figure}[t!]
		\centering
		\includegraphics[width=0.9\linewidth]{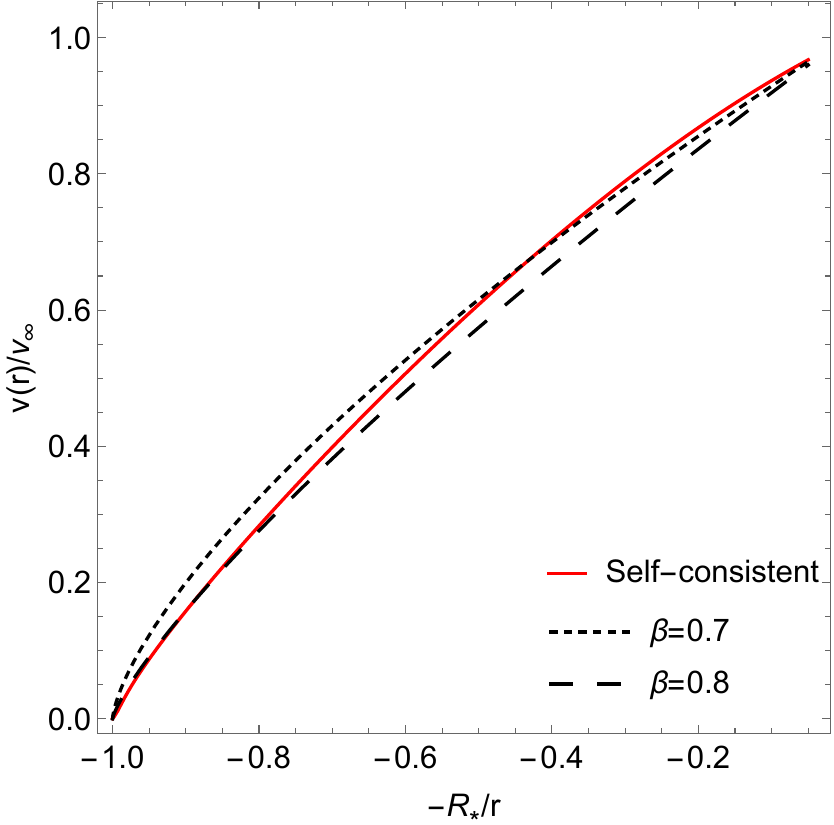}\\
		\vspace{5mm}
		\includegraphics[width=0.9\linewidth]{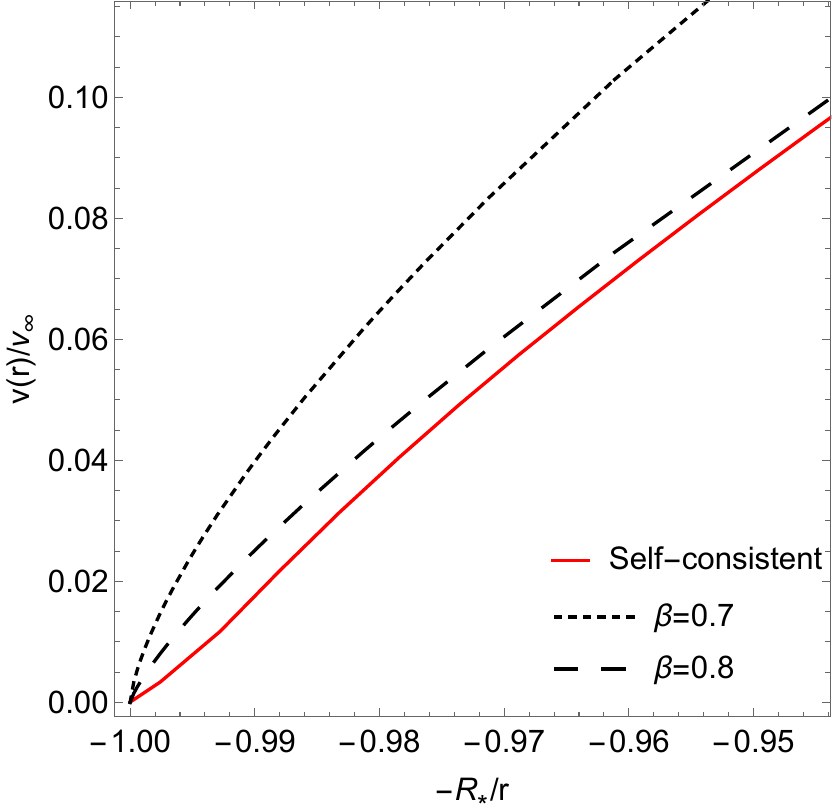}
		\caption{\small{Velocity profile generated from a self-consistent wind solution using the m-CAK prescription, compared with some $\varv(r)$ following $\beta$-law. The bottom panel corresponds to a zoom into the base of the wind.}}
		\label{newvel}
	\end{figure}

	As a consequence, new theoretical values for the wind parameters (mass-loss rate, terminal velocity) can be determined for any specific set of stellar parameters (effective temperature, surface gravity, stellar radius, abundances).
	In the particular case of mass-loss rates, self-consistent values have proven to agree with those determined by observations when homogeneous wind is assumed \citep[see figure~13 in][]{alex19}; whereas for the case of clumped winds, the clumping factor is a free parameter in the spectral fitting.
	This is an important result because these spectral fittings were achieved without using the power law that is commonly assumed to describe the velocity profile of stellar winds, namely $\beta$-law
	\begin{equation}
		\varv(r)=\varv_\infty\left(1-\frac{R_0}{r}\right)^\beta\;.
	\end{equation}
	This description, even when adjusted to the observations, is not derived from hydrodynamic calculations and then is justified a posteriori \citep{kudritzki00}.
	Therefore, self-consistent velocity fields cannot be described by any $\beta$ exponent (see Fig.~\ref{newvel}), and thus they need to be calculated from the m-CAK equation of motion (Eq.~\ref{motion1}) following the recipe introduced by \citet{michel04}.
	This departure from $\beta$-law agrees with that observed for the self-consistent solutions calculated from the Lambert-procedure in \citet{alex21a}, where the new velocity profile could not be recovered by any $\beta$ exponent.
	However, it is necessary to remark that the calculated velocity field differs from the line-force parameters, compared with hydro solutions in the CMF: if we compare our Figure~\ref{newvel} with Figure~9 from \citet{alex21a}, we observe a steeper profile at the base of the wind for the case of $\varv(r)$ calculated from m-CAK prescription.
	This is because the CMF approach accounts for non-Sobolev effects such as source-function dip, back-scattering, or line overlaps, and line-acceleration calculated in CMF additionally does not present an explicit dependence on velocity gradient or velocity \citep[see discussion from Section~5.3 of][]{sundqvist19}, and therefore it is very different from the CAK framework.
	Nevertheless, in spite of these discrepancies between the velocity profile for the subsonic region of the wind, they are not translated into the synthetic spectra.
	We performed a posteriori self-consistent solutions with a subsonic region based on the CMF approach \citep[following the coupling strategy introduced in][see their Section 2.1.1]{alex21a} and eventually found that the resulting differences in the synthetic spectra are just marginal.
	Hence, we continued without modifying the subsonic region for our self-consistent velocity profiles.

	The spectral fitting for the self-consistent solutions from the m-CAK prescription was performed with the radiative transfer code FASTWIND \citep{santolaya97,repolust04,puls05}, which has the option of using a velocity profile generated from a formal solution of the equation of motion as input instead of a classical $\beta$-law \citep{araya17}.
	Thus, it is possible to create synthetic spectra in which the wind parameters are not longer free, but depend on the stellar parameters. This reduces the number of free parameters.
	Based on this statement, we implement a method for fitting spectral observations in which the wind parameters are subordinate to the fit of the stellar parameters that generate them.

	The self-consistent m-CAK solutions provided by Paper I were shown to be reliable for a wide range of O stars.
	Studies aiming for a self-consistent prescription under a full NLTE treatment, such as \citet{alex21a}, have provided only slightly different results.
	However, we considered improvements of our m-CAK prescription necessary, especially to rectify the high terminal velocities provided by it.
	For this reason, we introduce in the next subsection a stratification for the radial temperature of the wind, which replaces the assumption of an isothermal wind for the calculation of the line-force parameters.

\subsection{Stratification of temperature}\label{stratification}
	\begin{figure*}%[t!]
		\centering
		\includegraphics[width=0.45\linewidth]{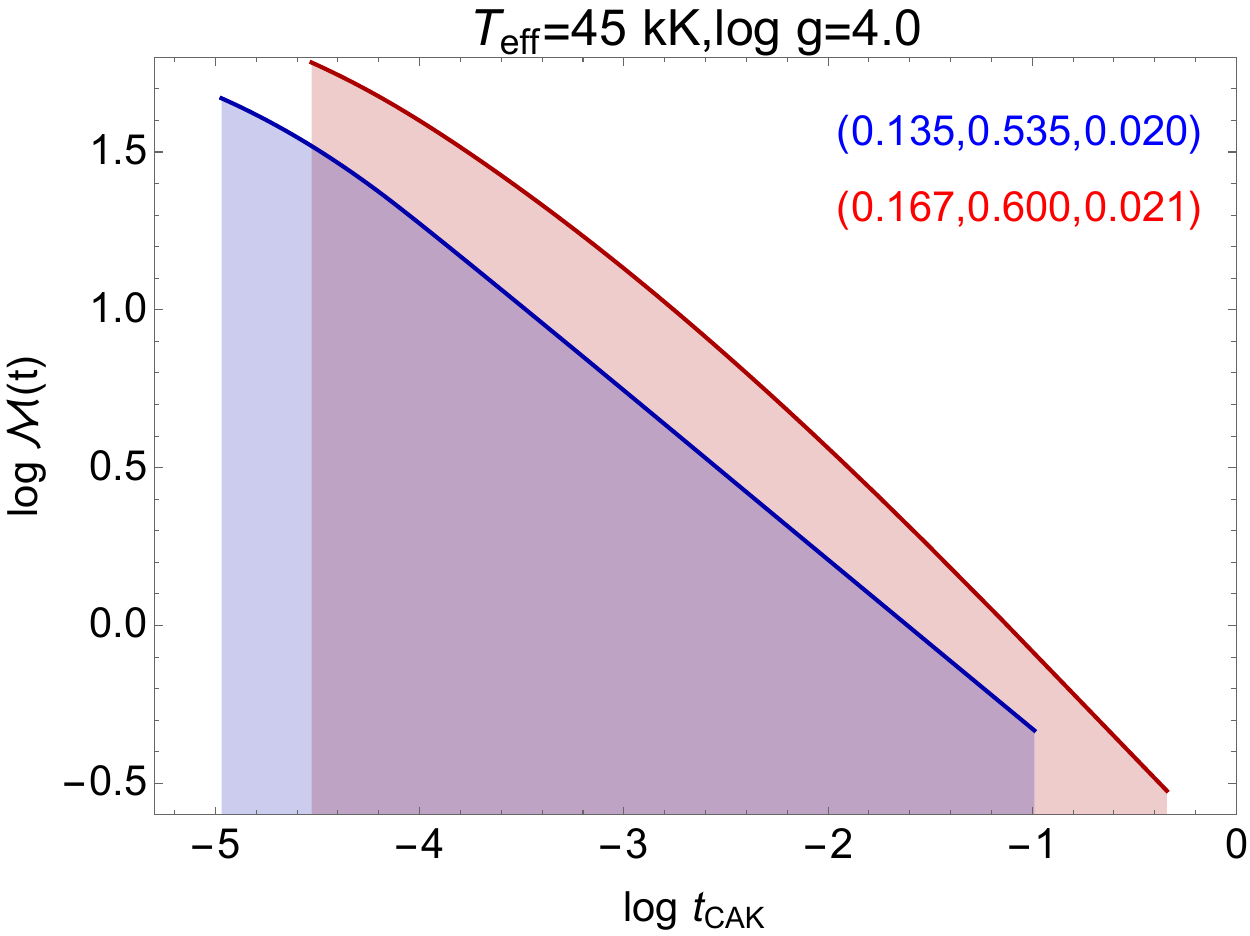}
		\includegraphics[width=0.45\linewidth]{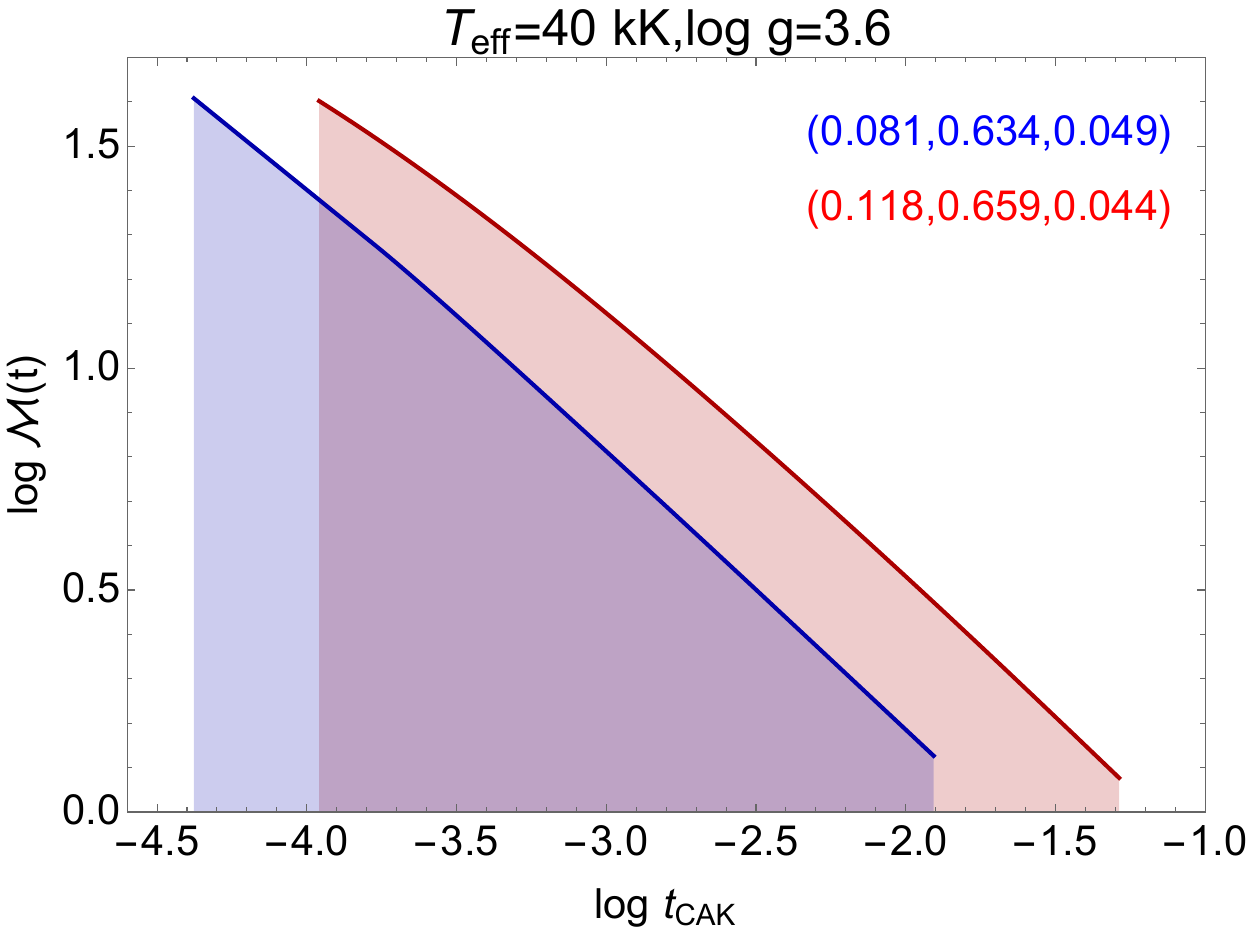}\\
		\includegraphics[width=0.45\linewidth]{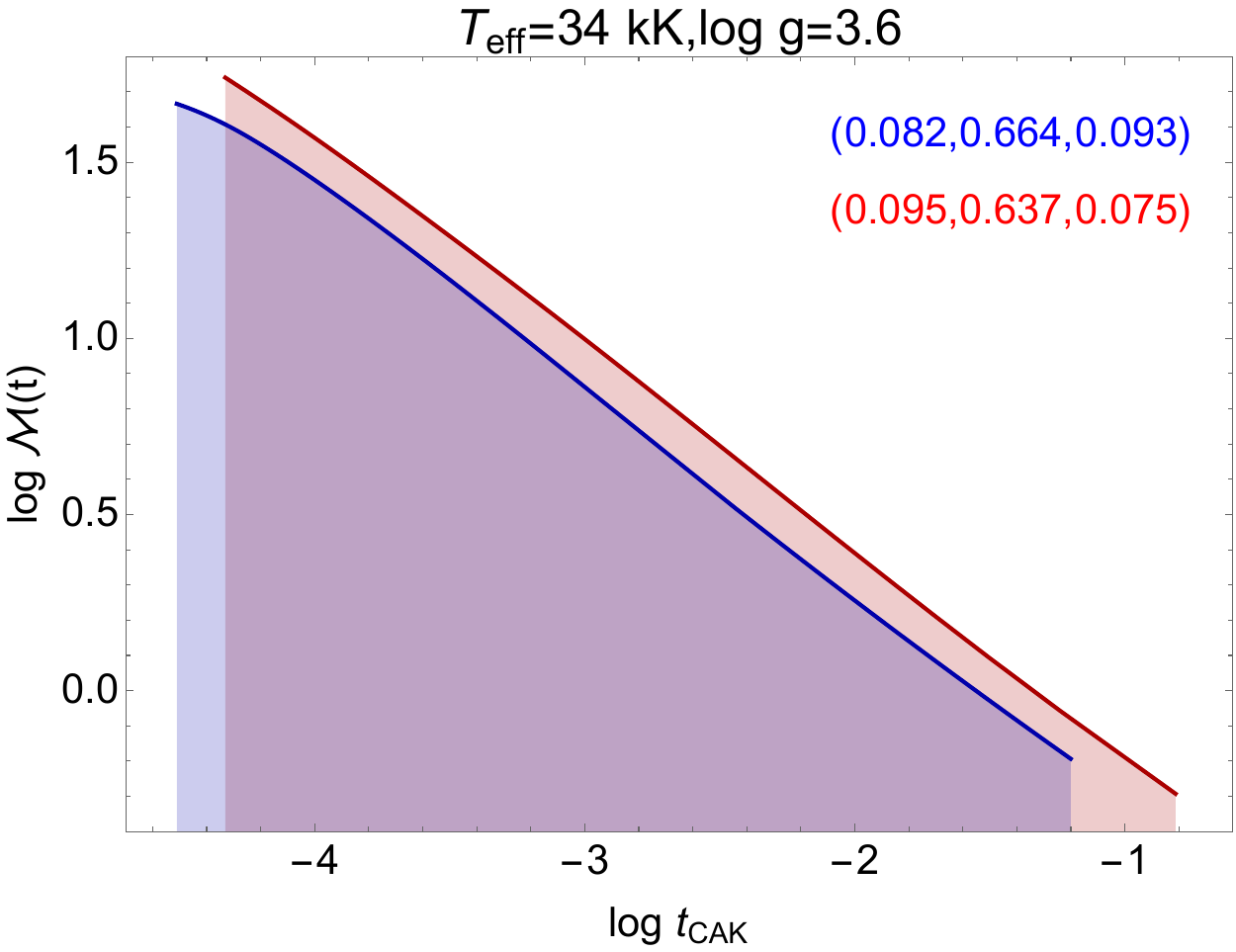}
		\includegraphics[width=0.45\linewidth]{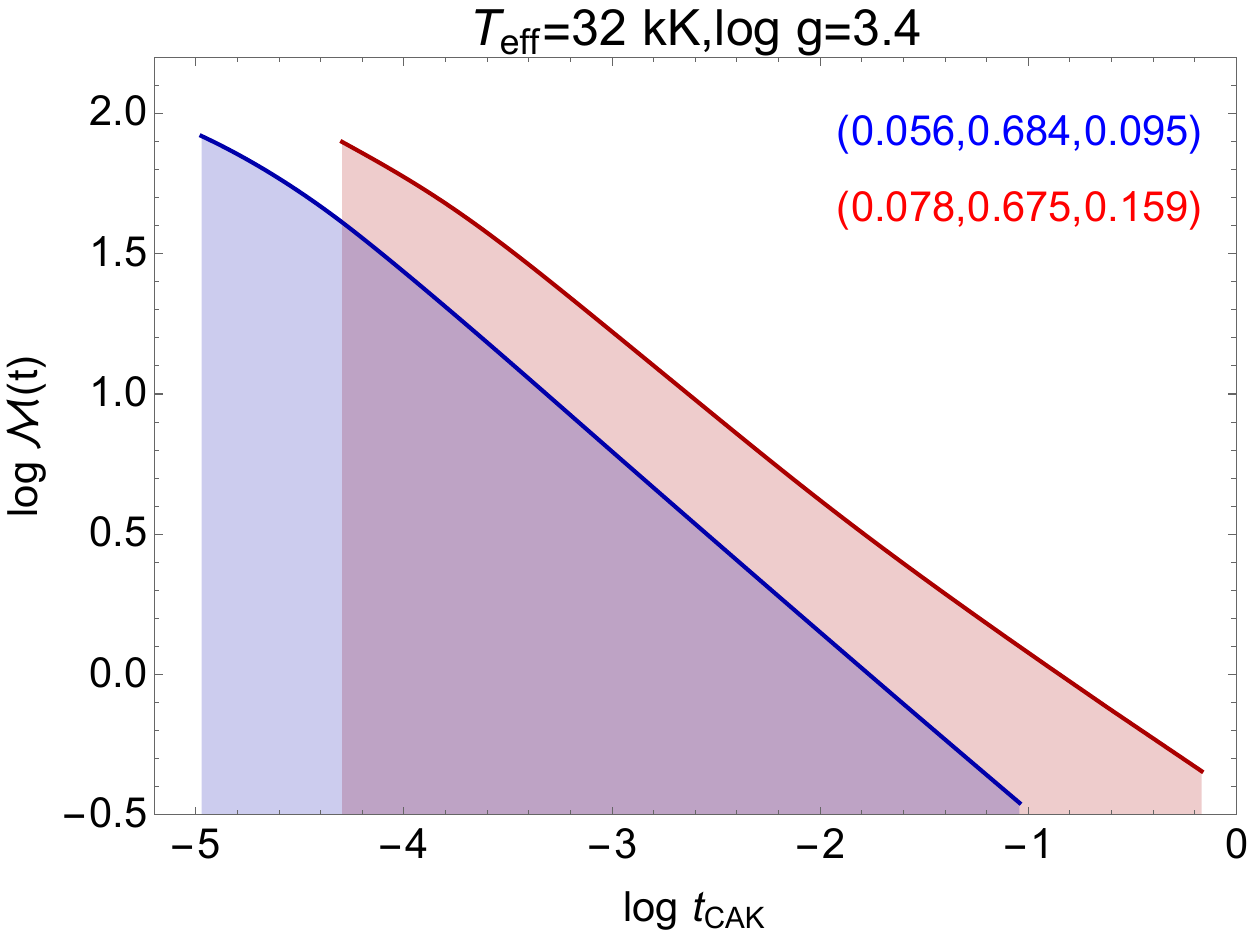}
		\caption{\small{Comparison of self-consistent fitting of the force multiplier $\mathcal M(t)$ between the implementation of the temperature structure (blue) and the fixed temperature (red). Calculated line-force parameters $(k,\alpha,\delta)$ are displayed in the same colour code. The shaded area represents the $t$ optical depth range where the force multiplier is defined within, from the sonic point outwards.}}
		\label{mtvslogt}
	\end{figure*}

	A complete inclusion of a temperature structure requires a sophisticated calculation of the radiative equilibrium and the electron balance through the wind, which is commonly made using radiative transfer codes such as CMFGEN, PoWR, or FASTWIND.
	Because the solution of radiative equilibrium is beyond the scope of our procedure, we introduce a stratification for the temperature field over the wind based on the simplified prescription used previously by \citet{sundqvist19}, and based on the \citet{lucy71} estimation for a temperature structure in a spherically extended envelope,
	\begin{equation}\label{trad}
		T(r)=T_\text{eff}\left[W(r)+\frac{3}{4}\tau_\text{F}\right]^{1/4}\;,
	\end{equation}
	with $W(r)$ being the dilution factor
	\begin{equation}\label{wdilution}
		W(r)=\frac{1}{2}\left(1-\sqrt{1-R_*^2/r^2}\right)\;,
	\end{equation}
	$\tau_\text{F}$ being the flux-weighted optical depth
	\begin{equation}\label{tauf}
		\tau_\text{F}(r)=\int\kappa_\text{F}(r')\,\rho(r')\left(\frac{R_*}{r'}\right)^2\,dr'\;,
	\end{equation}
	and $\kappa_\text{F}$ being the flux-weighted mean opacity
	\begin{equation}
		\kappa_\text{F}(r)F(r)=\int_0^\infty\kappa_\nu F_\nu d\nu\;.
	\end{equation}
	This means that the mean opacity, even when it typically is calculated by radiative transfer, is directly connected to the radiative acceleration of the stellar wind \citep[see Eq.~3 in][]{sundqvist19},
	\begin{equation}
		g_\text{rad}(r)=\frac{\kappa_\text{F}(r)L_*}{4\pi cr^2}=\frac{\kappa_\text{F}(r) F(r)}{c}\;,
	\end{equation}
	Then, including the relation between radiative and line-acceleration \citep[Eq.~3 and 6 from][]{alex21a}, it is possible to rewrite the flux-weighted optical depth as
	\begin{equation}\label{taufinal}
		\tau_\text{F}(r)=\frac{4\pi cR_*^2}{L_*}\int \left[g_\text{line}(r')+g_\text{grav}(r')\Gamma_\text{e}\right]\,\rho(r')\,dr'\;,
	\end{equation}
	with $g_\text{grav}(r)=GM_*/r^2$ and $\Gamma_\text{e}$ being the Eddington factor \citep[][Eq.~4]{alex21a}.
	For details, see the step-by-step calculation in Appendix~\ref{tau}.

	Because the entities inside the integral are calculated by our m-CAK prescription, it is possible  to compute the flux-weighted mean optical depth simultaneously with the iterative loop calculations, and thus the final stratificated temperature is self-consistent with the final line-acceleration.
	Like \citet{sundqvist19} and \citet{puls05}, we imposed a ground temperature of $T(r)=0.4\,T_\text{eff}$ to avoid excessively low temperatures at larger distances.

	\begin{table*}[t!]
		\centering
		\caption{\small{Values for self-consistent $(k,\alpha,\delta)$, $\dot M$, and $\varv(r)$ when the temperature structure is included.}}
		\begin{tabular}{cccccccc}
			\hline\hline
			$T_\text{eff}$ & $\log g$ & $k$ & $\alpha$ & $\delta$ & $\dot M$ & $\varv_\infty$ & $\log\dot M$\\
			(kK) & & & & & ($10^{-6}\,M_\odot$ yr$^{-1}$) & (km s$^{-1}$)\\
			\hline
			45 & 4.0 & $0.135\pm.007$ & $0.535\pm.005$ & $0.020\pm.002$ & $0.65\,(-68\%)$ & $2\,590\pm130\,(-25\%)$& $-6.187\pm.081$\\
			40 & 3.6 & $0.081\pm.003$ & $0.634\pm.004$ & $0.049\pm.002$ & $2.9\,(-56\%)$ & $2\,410\pm140\,(-14\%)$ & $-5.538\pm.053$\\
			34 & 3.6 & $0.082\pm.004$ & $0.664\pm.004$ & $0.093\pm.002$ & $1.3\,(+8\%)$ & $2\,790\pm160\,(+2\%)$ & $-5.886\pm.067$\\
			32 & 3.4 & $0.056\pm.002$ & $0.684\pm.003$ & $0.095\pm.001$ & $0.69\,(-43\%)$ & $2\,080\pm250\,(+26\%)$ & $-6.161\pm.051$\\
			\hline
		\end{tabular}
		\tablefoot{\small{Variations with respect to the self-consistent values tabulated by \citet[][Table~3]{alex19}, expressed in percentages for the wind parameters, are listed in the parentheses.
		For a detailed explanation of the error bars, see Section~\ref{errorbars}.}}
		\label{tabletemperatures}
	\end{table*}

%_____Temperature structure and line-force parameters_____________
\subsection{Temperature structure and line-force parameters}\label{tradimplications}
	The direct effects of the change in temperature over the wind are the alteration of the excitation and ionisation stages for the atomic populations, being the lower levels more present in the more external parts of the wind with respect to the case of a fixed temperature for the wind.
	Since lower stages of ionisation tend to have more lines \citep{abbott82,puls00}, a temperature structure should generate a larger contribution to the line force in the outer region of the wind, which can be translated into higher values for the line-force parameter $\alpha$.
	However, this trend is observed only for low values of $T_\text{eff}$: variation in $\alpha$ passes from diminishing its value (in comparison with the fixed-temperature case) for hotter effective temperatures to slightly enhance it for cooler $T_\text{eff}$.
	At the same time, the variation in $\delta$ (directly linked with the ionisation of the wind) seems to be just marginal, with the exception of the 32 kK star.

	Another consequence of the stratification of the temperature is the restructuring of the optical depth $t$ (Eq.~\ref{t}), which is implicitly dependent on $T(r)$, in particular, $\varv_\text{th}$, which is the mean thermal velocity of the protons of the wind,
	\begin{equation}\label{vtherm}
		\varv_\text{th}(r)=\sqrt{\frac{2k_\text{B}T(r)}{m_\text{H}}}\;,
	\end{equation}
	with $k_\text{B}$ being the Boltzmann constant.
	Because the temperature through the wind is no longer constant, this thermal velocity decreases as radial distance increases, and subsequently, the optical depth.
	As a result, we observe from Fig.~\ref{mtvslogt} that the range in which the force multiplier is calculated appears to be shifted to lower values for $t$.
	Subsequently, this displacement produces a direct decrement of the line-force parameter $k$, and therefore also of $\alpha$ and $\delta$ due to the iterative loop searching for self-consistency.

	The wind parameters calculated from our four analysed standard stars are tabulated in Table~\ref{tabletemperatures}.
	When the temperature structure is included, the mass-loss rate is clearly most strongly affected: from a great decrease at high temperatures (45 kK) to a marginal gain at cooler temperatures (34 kK), and another high declination for the case $T_\text{eff}=32$ kK (which can altogether be explained by the reduction in the line-force parameter $\delta$).
	For modifications of the terminal velocities are directly linked with the correction of the values for $\alpha$, and they are again due to the reduction of $\delta$ for the coolest case.

	\begin{figure*}[htbp]
		\centering
		\includegraphics[width=0.45\linewidth]{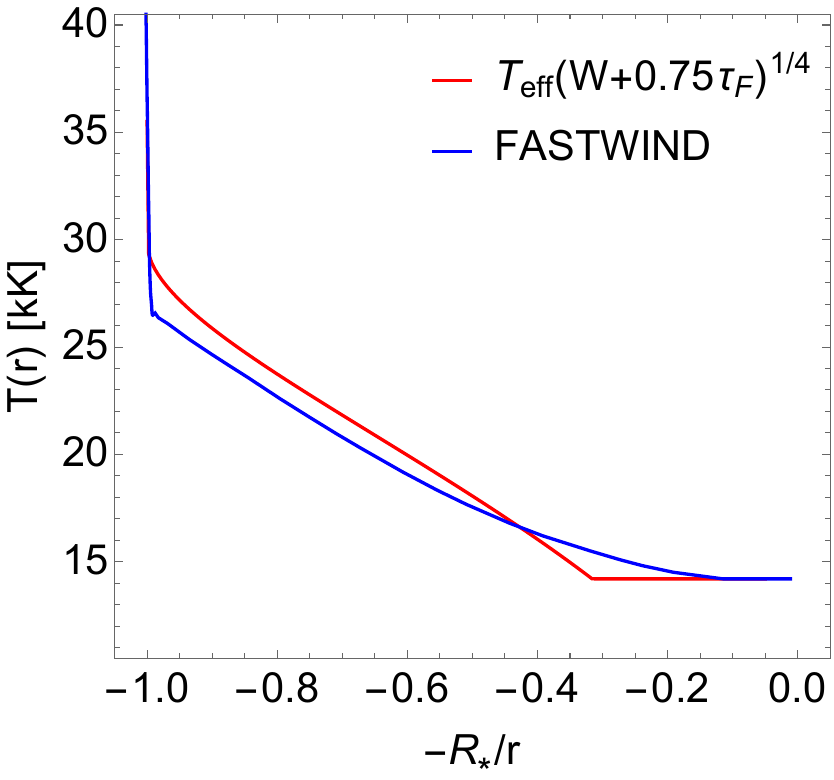}
		\includegraphics[width=0.45\linewidth]{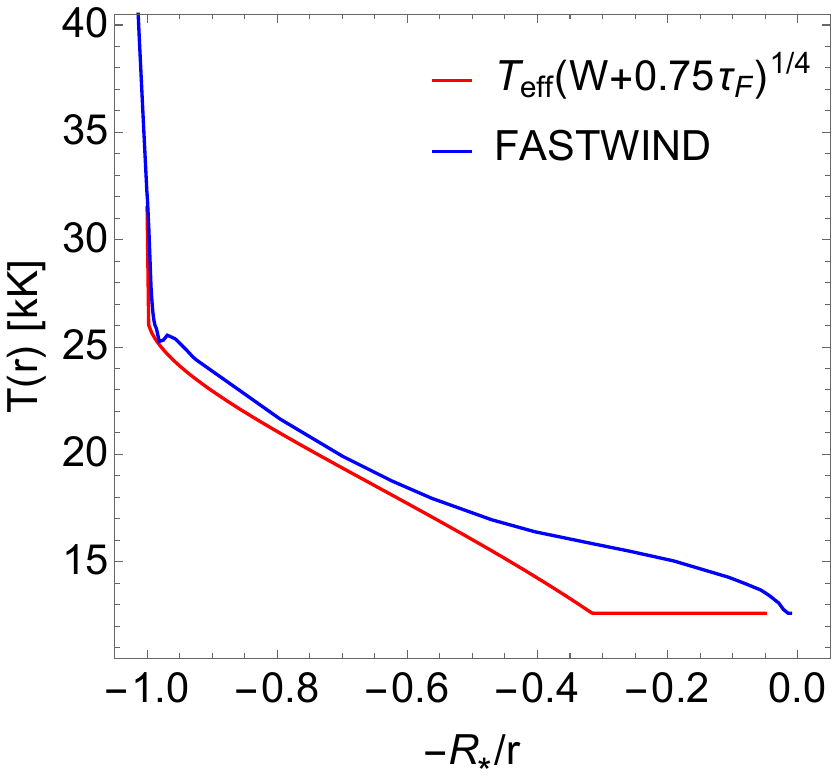}
		\caption{\small{Temperature stratification for HD 57682 (left panel, $T_\text{eff}=35$ kK) and HD 210809 (right panel, $T_\text{eff}=31.5$ kK), either obtained from Eq.~\ref{trad} (red) or calculated by FASTWIND (blue).}}
		\label{trad_fastwind}
	\end{figure*}

	This procedure of calculating a stratified temperature was made under the assumption that we can recover opacities and emissivities from our m-CAK line-acceleration.
	In other words, our implemented $T(r)$ is not calculated by radiative transfer and therefore will differ from a fully calculated temperature structure.
	Because our synthetic spectra are determined by FASTWIND, we can compare the stratificated temperature that was used during the iterative procedure calculating our self-consistent parameters and the temperature obtained with opacities and emissivities calculated by radiative transfer.
	These comparisons for two stars of our sample are shown in Fig.~\ref{trad_fastwind}, where we observe that differences in $T(r)$ are $\sim3$ kK (except for the external part of HD 210809).
	It might be argued that the similarity in the results is forced because the two prescriptions use the same boundary conditions (where $T(r)=T_\text{eff}$ for the photosphere and $T(r)=0.4\,T_\text{eff}$ for the imposed external limit), but even in this case, the replacement of our m-CAK temperature structure by the fully calculated by FASTWIND produces only marginal variations in the self-consistent parameters with a higher computational effort.
	Therefore, the prescription for the stratification of temperature introduced in this work, which is considered as a general approximation, quickly leads to more reliable results than the case $T(r)=T_\text{eff}$ for our self-consistent wind parameters.

	\begin{table*}[t!]
		\centering
		\caption{\small{Reformulation of the grid of wind solutions from KK17 (their Table~1), using their stellar parameters, but performing our own self-consistent solutions for the line-force and wind parameters.}}
		\resizebox{\textwidth}{!}{
		\begin{tabular}{cccc|ccc|ccrr}
			\hline\hline
			Model & $T_\text{eff}$ & $\log g$ & $R_*$& $k$ & $\alpha$ & $\delta$ & $\varv_\infty$ & $\log\dot M_\text{sc}$ & $\log\left(\frac{\dot M_\text{sc}}{\dot M_\text{Vink}}\right)$ & $\log\left(\frac{\dot M_\text{sc}}{\dot M_\text{KK17}}\right)$\\
			& $[\text{K}]$ & & $[R_\odot]$ & & & & [km s$^{-1}$] & [$M_\odot\,\text{yr}^{-1}$]\\
			\hline
			\multicolumn{4}{c}{Main sequence (V)} & \\
			325-V & 32.5 & 3.92 & 7.4 & $0.015\pm.001$ & $0.672\pm.005$ & $0.032\pm.001$ & $3\,330\pm180$ & $-8.222\pm.097$ & $-0.855$ & $-0.368$\\
			350-V & 35.0 & 3.92 & 8.3 & $0.131\pm.006$ & $0.607\pm.004$ & $0.090\pm.002$ & $2\,210\pm120$ & $-6.783\pm.096$ & $0.137$ & $0.636$\\
			375-V & 37.5 & 3.92 & 9.4 & $0.169\pm.010$ & $0.524\pm.005$ & $0.046\pm.002$ & $1\,980\pm130$ & $-6.816\pm.101$ & $-0.293$ & $0.143$\\
			400-V & 40.0 & 3.92 & 10.7 & $0.162\pm.009$ & $0.495\pm.006$ & $0.015\pm.003$ & $2\,040\pm130$ & $-6.732\pm.095$ & $-0.595$ & $-0.130$\\
			425-V & 42.5 & 3.92 & 12.2 & $0.136\pm.007$ & $0.519\pm.005$ & $0.013\pm.002$ & $2\,310\pm130$ & $-6.380\pm.090$ & $-0.584$ & $-0.034$\\
			\hline
			\multicolumn{4}{c}{Giants (III)} & \\ 
			300-III & 30.0 & 3.49 & 13.1 & $0.048\pm.001$ & $0.684\pm.002$ & $0.102\pm.001$ & $2\,070\pm220$ & $-6.789\pm.111$ & $0.007$ & $0.419$\\
			325-III & 32.5 & 3.53 & 13.4 & $0.065\pm.002$ & $0.680\pm.003$ & $0.100\pm.001$ & $2\,100\pm180$ & $-6.385\pm.068$ & $0.047$ & $0.536$\\
			350-III & 35.0 & 3.58 & 13.9 & $0.081\pm.003$ & $0.661\pm.004$ & $0.085\pm.002$ & $2\,150\pm140$ & $-6.111\pm.066$ & $-0.014$ & $0.509$\\
			375-III & 37.5 & 3.63 & 14.4 & $0.092\pm.004$ & $0.637\pm.005$ & $0.068\pm.002$ & $2\,190\pm140$ & $-5.943\pm.070$ & $-0.120$ & $0.433$\\
			400-III & 40.0 & 3.67 & 15.0 & $0.093\pm.004$ & $0.615\pm.005$ & $0.049\pm.002$ & $2\,170\pm130$ & $-5.858\pm.069$ & $-0.238$ & $0.152$\\
			425-III & 42.5 & 3.72 & 15.6 & $0.091\pm.003$ & $0.595\pm.004$ & $0.031\pm.002$ & $2\,220\pm130$ & $-5.810\pm.060$ & $-0.463$ & $0.014$\\
			\hline
			\multicolumn{4}{c}{Supergiants (I)} & \\
			325-I & 32.5 & 3.30 & 21.4 & $0.054\pm.002$ & $0.689\pm.003$ & $0.092\pm.002$ & $1\,910\pm220$ & $-5.887\pm.081$ & $0.120$ & $0.463$\\
			350-I & 35.0 & 3.41 & 20.5 & $0.065\pm.002$ & $0.674\pm.004$ & $0.077\pm.002$ & $2\,100\pm200$ & $-5.732\pm.068$ & $0.038$ & $0.512$\\
			375-I & 37.5 & 3.52 & 19.8 & $0.077\pm.003$ & $0.657\pm.004$ & $0.065\pm.002$ & $2\,280\pm160$ & $-5.606\pm.067$ & $-0.174$ & $0.315$\\
			400-I & 40.0 & 3.63 & 19.1 & $0.084\pm.003$ & $0.627\pm.004$ & $0.046\pm.002$ & $2\,430\pm140$ & $-5.629\pm.067$ & $-0.301$ & $0.115$\\
			425-I & 42.5 & 3.75 & 18.5 & $0.096\pm.004$ & $0.573\pm.005$ & $0.023\pm.002$ & $2\,470\pm130$ & $-5.779\pm.068$ & $-0.527$ & $-0.034$\\
			\hline
		\end{tabular}}
		\tablefoot{\small{Differences in logarithm scale on mass-loss rate, compared with both formulae from \citet{vink01} and \citet{kk17}, are included in the last two columns (see also Fig.~\ref{mdotsvinkkk17}).}}
		\label{tablekk17}
	\end{table*}

%_____Variance of line-force parameters
\subsection{Variance of line-force parameters}\label{errorbars}
	The range of validity for the self-consistent solutions from Paper I was established for effective temperatures from 32 to 45 kK and for surface gravities with $\log g\ge3.4$ because within these limits, the uncertainties over the final wind parameters resulting from the fitting of the line-force parameters were smaller than the error bars derived from the observational uncertainties of the stellar parameters.
	This was made on the basis of a qualitative analysis of the behaviour of the force multipliers $\mathcal M(t)$.
	We aim to complement this outline by including a quantitative measurement of the fit for the line-force parameters by means of calculating their respective standard deviations.
	Hereafter, error bars associated with the wind parameter mass-loss rate and terminal velocity come from the variation in $(k,\alpha,\delta)$ instead of from the variation in stellar parameters.

	It is important to recall that these analyses over the error bars come from the assumption that line-force parameters are constant throughout the wind instead of a function of depth, as has been proposed by authors such as \citet{schaerer94} and \citet{kudritzki02}.
	In other words, we can consider $k$, $\alpha,$ and $\delta$ as constants only if their uncertainties lies below the uncertainties found for the stellar parameters. This condition is satisfied for the range of temperatures we studied, where $\log\mathcal M(t)$ presents an almost linear shape as a function of $\log t$ (see Fig.~\ref{mtvslogt}).

	From Paper I, the uncertainties on the stellar parameters generated errors of $\sim7-15\%$ for terminal velocities and of $\sim20\%$ for mass-loss rates ($\sim0.1$ in logarithmical scale).
	Hence, we could consider these values the threshold of the maximum permisible error on our models, with special emphasis on $\dot M$.
	For this reason as well, the values for the mass-loss rate and their uncertainties are always given in logarithmical scale here.
	
	The line-force parameters introduced in Table~\ref{tabletemperatures} are given with errors, which come from the standard deviation for each parameter when the fit is performed.
	Likewise, the terminal velocities and the logarithm of the mass-loss rates contain error bars that represent the new associated uncertainties for our self-consistent wind parameters.
	These uncertainties do not come only from the standard deviation of $(k,\alpha,\delta)$, but also from the discrepancies produced by numerical issues such as the selected number of depth points when the equation of momentum (hereafter e.o.m., Eq.~\ref{motion1}) is calculated by \textsc{HydWind} \citep[for details of the numerical calculation of the e.o.m., see][]{michel04}.
	This is calculated by summing them quadratically,

	\begin{equation}\label{deltawind}
		\Delta_\text{wind}=\sqrt{\Delta_{(k,\alpha,\delta)}^2+\Delta_{N_\text{points}}^2}\;,
	\end{equation}
	with $\Delta_{(k,\alpha,\delta)}$ and $\Delta_{N_\text{points}}$ being the error intervals either for $\varv_\infty$ or $\log\dot M$ generated by the variation in line-force parameters and the number of points, respectively.

	\begin{figure*}[t!]
		\centering
		\includegraphics[width=0.45\linewidth]{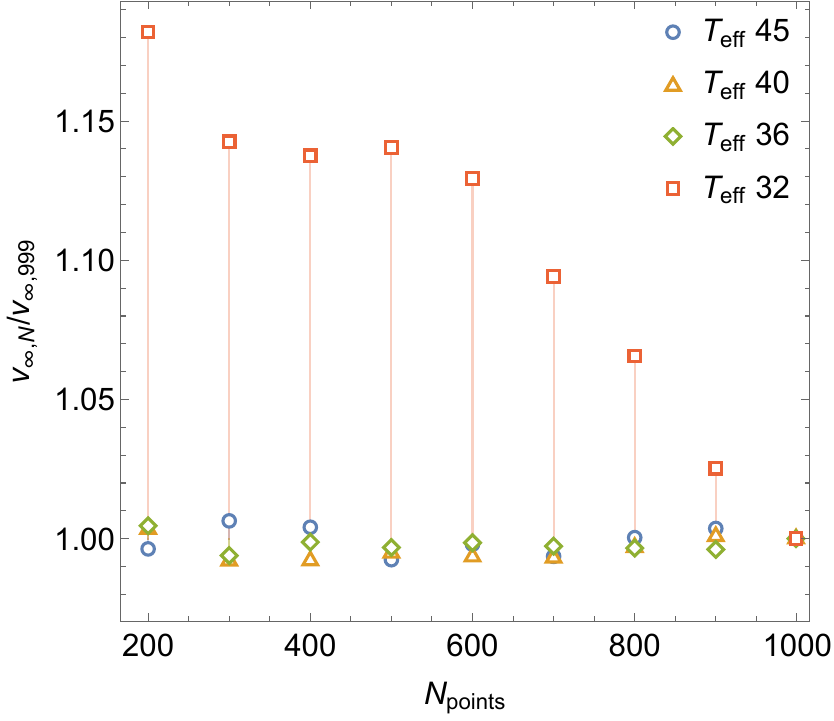}
		\includegraphics[width=0.45\linewidth]{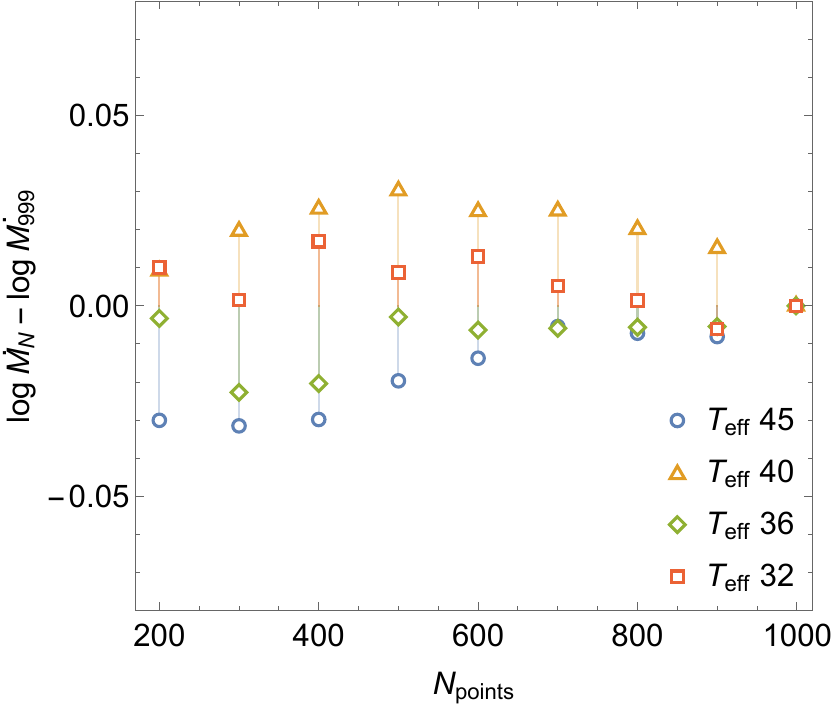}
		\caption{\small{Comparison of the obtained self-consistent wind parameters $\varv_\infty$ and $\dot M$, employing different numbers of depth points $N$ to numerically solve the equation of motion. Values are expressed as departures from the self-consistent terminal velocity and mass-loss rate when the maximum number of points is allowed for \textsc{HydWind} (999).}}
		\label{npoints_wind}
	\end{figure*}

	For our selected four standard stars, self-consistent wind parameters were calculated for a set of different numbers of depth points from 200 to 900 (in intervals of 100), plus the maximum number of points allowed by the code (999).
	An increase in the number of depth points results in a fine grid for the wind structure, but implies higher time consumption.
	By normalising these values by the terminal velocity and mass-loss rate under the maximum $N$ ($\varv_{\infty,999}$ and $\dot M_{999}$), the deviation generated for the use of different number of points is illustrated better, as is shown in Fig.~\ref{npoints_wind}.
	It is observed that only marginal discrepancies are created for mass-loss rates of $\sim0.03$ in logarithmical scale (below the threshold of 0.1 introduced above), whereas a remarkable deviation for the terminal velocity appears for the star with $T_\text{eff}=32$ kK and $\log g=3.4$.
	Therefore, uncertainties over $\varv_\infty$ increase considerably for lower values of the effective temperature and surface gravity, which reach $\sim12-13\%,$ as is shown in Table~\ref{tabletemperatures}, near to the $\sim15\%$ previously measured from the variation in stellar parameters.
	However, we emphasise that these are partial trends, and more models are needed to generate a definitive conclusion.

	Including these error intervals allows us a more quantitative analysis of the validity range of  our self-consistent m-CAK prescription.
	This gives us the chance to explore and evaluate the existence of wind solutions beyond the rigid thresholds derived in Paper I.
	Throughout this work, we scrutinise the presence or absence of solutions for effective temperatures cooler than 32 kK and surface gravities below 3.4.
	\begin{figure}[t!]
		\centering
		\includegraphics[width=0.9\linewidth]{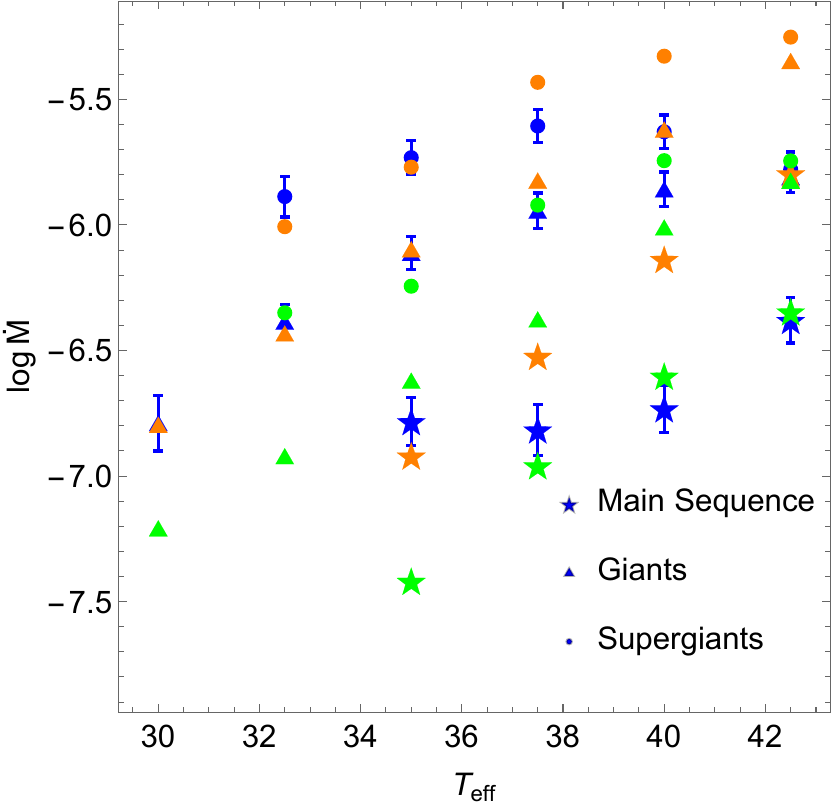}
		\caption{\small{Theoretical mass-loss rates tabulated in Table~\ref{tablekk17} for the values predicted by Vink's formula (orange symbols), by KK17 (green symbols), and by this study (blue symbols). Main-sequence stars of the sample are represented by starss, giants by triangles, and supergiants by circles.}}
		\label{mdotsvinkkk17}
	\end{figure}

	\begin{table*}[t!]
		\centering
		\caption{\small{Mercator-HERMES observations.}}
		\begin{tabular}{lllcccr}
			\hline\hline
			Star & Oth. name & Spec. type & $V$ (mag) & No. exp. & Exp. time (s) & Obs. date \\
			\hline
			HD 57682  & HR 2806       & O9.2 IV  & 6.43 & 5 & 1800 & 20 Nov 2012 \\
			HD 192639 & BD$+$36 3958  & O7.5 Iab & 7.11 & 1 & 800  & 02 Jul 2013 \\
			HD 188001 & 9 Sge	 & O7.5 Iab & 6.23 & 3 & 313  & 30 Sep 2013 \\
			HD 218915 & BD$+$52 3383  & O9.2 Iab  & 7.20 & 1 & 2180 & 20 Aug 2013 \\
			HD 210809 & BD$+$51 3281  & O9 Iab   & 7.56 & 1 & 1200 & 03 Nov 2013 \\
			HD 195592 & MWC 347       & O9.7 Ia  & 7.08 & 1 & 1800 & 23 Aug 2013 \\
			\hline
		\end{tabular}
		\label{hermesobservations}
	\end{table*} 

%_____Range of validity
\subsection{Range of validity}\label{temperaturerange}
	Before we implement the synthetic spectra to fit observed spectra, a quick evaluation of our self-consistent wind parameters with respect to previous studies is necessary.
	Analogously to Paper I, we compared our resulting self-consistent mass-loss rates, $\dot M_\text{sc}$ with the theoretical values provided by the Monte Calo procedure of \citet{vink00,vink01}, and for this case, we specifically include a comparison with the self-consistent wind solutions for O-type stars calculated by KK17.
	We selected these two studies because they present a surprisingly well-correlated theoretical mass-loss rate, with the values from KK17 being $\text{about three}$ times lower (a difference of $\sim0.5$ in logarithmical scale) than those given by the so-called Vink formulae.

	For this purpose, we took the stellar parameters from the sample introduced by KK17 and calculated our own self-consistent wind solutions.
	The results are displayed in Table~\ref{tablekk17}, and we compare the different obtained mass-loss rates in Fig.~\ref{mdotsvinkkk17}.
	Two stars with effective temperatures of 30 kK, 300-V and 300-I (see Table~1 from KK17), were not included in this comparison.
	The main-sequence star (300-V) was excluded because the values of stellar radius ($6.6\,R_\odot$) and mass ($12.9\,M_\odot$) are too to solve the e.o.m. by \textsc{HydWind}, and the supergiant (300-I) was excluded because the low temperature and surface gravity led to an unstable model without a unique self-consistent solution (for details, see Section~\ref{multipleselfconsistency}).
	For the remaining stars, it is interesting to note that high effective temperatures generate values for $\dot M_\text{sc}$ that agree more closely with those provided by KK17, whereas (with the remarkable exception of 325-V, the outlier of our sample) mass-loss rates from cooler temperatures show a better correspondence with Vink's formulae.
	This unexpected trend is curious because first, our theoretical self-consistent mass-loss rates differ from the good correlation observed between $\dot M_\text{Vink}$ and $\dot M_\text{KK17}$, and second, it would be possible to establish a smooth transition for theoretical $\dot M$ at some point around $T_\text{eff}~32.5$ kK and $\log g~3.5-3.3$ for future evolutionary tracks of massive stars if we were to change the mass-loss prescription from self-consistent values to Vink's formulae \citep[][in prep.]{alex22}.

%_____OBSERVATIONAL DATA__________________________________________________________
\section{Observational data}\label{hermesdata}

	The selected sample is composed of a set of O supergiants (HD 192639, 9 Sge, HD 218915, HD 210809 and HD 195592), which are situated at the lower border of validity for effective temperature and surface gravity as introduced in Section~\ref{temperaturerange}.
	Additionally, we included a star known by its magnetic field (HD 57682) in order to confirm the quality of a spectral fitting for wind structures departing from spherically symmetry.
	For the spectral data, the sample was chosen based on the high quality of the signal-to-noise ratio (S/N) and for stars with compelling evidence that they are single stars (i.e. excluding any possible binarity).

%_____Hermes spectra
\subsection{\textsc{Hermes} spectra}

	The spectra were observed with the \textsc{Hermes} echelle spectrograph mounted on the Mercator 1.2m telescope at the Roche de los Muchachos Observatory, La Palma, Spain.
	\textsc{Hermes} (High Efficiency and Resolution Mercator Echelle Spectrograph) is a high-efficiency prism-cross-dispersed fibre-fed bench-mounted spectrograph that observes the complete wavelength range $3800-9000\,\AA$ in a single exposure at a spectral resolution of $R~=~85\,000$ \citep{raskin11}.

	Each star was observed using the high-resolution fibre mode in one to five exposures, depending on the $V$ magnitude, in succession throughout the same night (see Table~\ref{hermesobservations}).
	Additional flat-field exposures were observed during the night to avoid introducing systematic noise in the flux calibration process.
	The spectra were calibrated with the latest version of the \textsc{Hermes} pipeline (release v6.0) developed at the Royal Observatory of Belgium and the University of Leuven, in collaboration with the \textsc{Hermes} Consortium.
	The typical calibration steps were performed, including spectral order tracing and extraction, average flat-fielding, Th-Ar lamp wavelength calibration, and cosmic ray removal using cross-order profiling.

	The calibrated exposures were co-added to produce a single stacked 1D spectrum per star with an S/N of $\sim150-250$.
	The S/N was estimated at four effective wavelengths ($U$, $B$, $V$, and $R$ bands).
	The typical uncertainty of the spectrograph wavelength calibration is below $\pm0.02\,\AA$, or the \textsc{Hermes} wavelength scale has a high accuracy.

%_____Continuum normalisation
\subsection{Continuum normalisation}

	The raw exposures were calibrated with the removal of cosmic rays for the final sum spectra.
	Dedicated software tools  were developed for BRASS \citep{lobel19} to remove cosmic ray flux
spikes in \textsc{Hermes} spectra.
	Moreover, the spectral response function of the instrument shows a strong wavelength dependence and some time variability.
	However, spectra observed with regular and smooth continuum flux shapes are required to perform automated continuum normalisation calculations.
	Consequently, all signatures of the instrumental response were removed from the observed spectra.
	Advanced \textsc{Hermes} calibration pipeline algorithms were previously developed in BRASS to determine the response correction curve at every point in time.

	The method for modelling the instrumental response curve is important to perform high-quality continuum flux normalisations, including in the small wavelength regions around selected spectral lines.
	We automatically normalised the observed spectra to the stellar continuum flux level using a special template normalisation procedure.
	It searches for variable wavelength points over sufficiently continuous flux regions close to the continuum level in theoretical (template) spectra to fold the \textsc{Hermes} spectra to these (continuum anchor) points.
	This ensures that local flux normalisation effects are minimised, providing reliably normalised spectra around the H and He lines we select for our detailed line profile modelling.

%_____METHOD_____________________________________________________________________________
\section{Method}\label{methodology}

%_____Selection of FASTWIND spectral lines
\subsection{Selection of FASTWIND spectral lines}
	The group of 15 lines calculated by FASTWIND is presented in Table~\ref{fastwindlines}.
	\begin{table}[h!]
		\centering
		\caption{\small{Lines included in FASTWIND, expressed in $\AA$.}}
		\begin{tabular}{lll}
			\hline\hline
			\multicolumn{1}{c}{Hydrogen} & \multicolumn{2}{c}{Helium}\\
			\hline
			H$\alpha$ 6563 & He II 4200 & He I 4387\\
			H$\beta$ 4861 & He II 4541 & He I 4471\\
			H$\gamma$ 4340 & He II 4686 & He I 4713\\
			H$\delta$ 4101 & He II 6527 & He I 4922\\
			H$\epsilon$ 3970 & He II 6683 & He I 6678\\
			\hline
		\end{tabular}
		\label{fastwindlines}
	\end{table}

	As a first stage, we decided to use only H and He lines for the spectral fitting in the optical and infrared wavelength range for practical purposes, whereas our observational data consist of optical spectra alone.

\subsection{Self-consistent modification of parameters}\label{modificationparameters}
	\begin{figure*}[t!]
		\centering
		\includegraphics[width=0.6\linewidth]{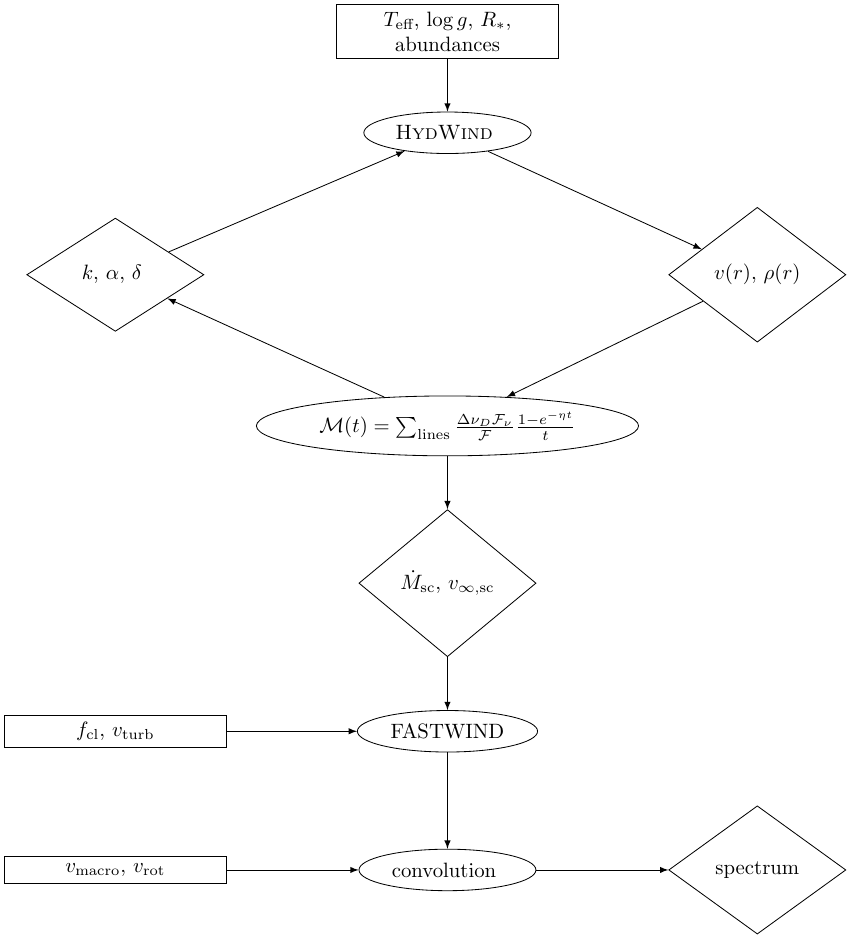}
		\caption{\small{Scheme of the self-consistent m-CAK procedure combined with FASTWIND, showing the stage at which the different initial parameters (in rectangles) are set.
		Ellipses represent the processes or codes to be run, and the diamonds represent the outputs.}}
		\label{mcakfastwind_flowchart}
	\end{figure*}

	The scheme showing the steps required to achieve a self-consistent synthetic spectrum is shown in Fig.~\ref{mcakfastwind_flowchart}.
	This procedure is the same as we used to obtain the spectra for Paper I, but now we go one step further: the spectral fit is improved by the fine-tuning of the free parameters (represented as rectangles in the Figure) at the different stages of the process.

	The upper half of Fig.~\ref{mcakfastwind_flowchart} corresponds to the iterative process that searches for the self-consistent solution under our m-CAK prescription, therefore its only input parameters are the stellar parameters (temperature, gravity, and abundances).
	When this solution is achieved, that is, when line-force parameters converge and theoretical values for the mass-loss rate, terminal velocity, and velocity profile are obtained, a synthetic spectrum is computed by FASTWIND using the new velocity profile $\varv(r)$ provided by the solution of $(k,\alpha,\delta)$ as an input.
	As we previously stated in Paper I, a clumping description is introduced at this point, together with the microturbulence $\varv_\text{turb}$.
	By clumping description we mean not only $f_\text{cl}$ , but also the starting and ending zones ($r_\text{in}$ and $r_\text{out}$) in which the inhomogeneities are located.
	This means that we use a clumping law in which $f_\text{cl}$ is constant over a region of the wind that is delimited by $r_\text{in}$ and $r_\text{out}$ \citep[see][for details]{puls06}.
	Finally, convolution for the resulting synthetic spectrum requires inserting a value for rotational velocity $\varv_\text{rot}$ and for macroturbulence $\varv_\text{macro}$ as input. 
	We therefore observe that input parameters can be classified as first-order (stellar parameters, whose modification implies a full restructuring of the self-consistent wind solution), second-order (clumping and turbulence, to recalculate the FASTWIND atmospheric model but without modifying the self-consistent wind solution), and third-order parameters (rotation and macroturbulence, to reshape line profiles).
	
	Although one of the advantages of developing synthetic spectra derived from self-consistent wind prescription is that the number of free parameters is reduced, the number of remaining parameters that need to be fit freely is still large enough to require a deeper evaluation of their effects upon the spectral lines.
	After many empirical tests and comparisons in which only one parameter was let free to vary while the others were held fixed, combined with previous analyses found in the literature, we observed some general trends that helped us to determine the best model fit.
	We describe them below.
	\begin{itemize}
		\item The effective temperature $T_\text{eff}$ is fitted by the ratio between the He II and the He I, given the expected change in ionisation.
		Because it is a first-order parameter, modifications of the effective temperatures imply an alteration of the self-consistent mass-loss rate (which are directly dependent), so that an increase or decrease in $T_\text{eff}$ will also affect the intensity of the hydrogen lines, especially H$\alpha$.
		Moreover, the effective temperature also provides information about the stellar radius, which is constrained with the luminosity (i.e. observed magnitude) by means of the Stefan-Boltzmann law.
		Therefore, modifications in $T_\text{eff}$ require modifications in $R_*/R_\odot$ for the models to maintain coherence with the $V$ magnitudes tabulated in Table~\ref{hermesobservations}.
		\item The surface gravity $\log g$ is evaluated by the absorption intensity observed for H$\beta$, H$\gamma$, H$\delta,$ and H$\epsilon$, together with the absorption profiles of the He I lines.
		Because it also is a first-order parameter, gravity presents an inverse proportionality with the self-consistent mass-loss rate.
		As a consequence, its increase implies a decrease in the potential emission component of H$\alpha$.
		\item The turbulent velocity $\varv_\text{turb}$, also known as microturbulence velocity, is fitted to the H I lines.
		In concordance with \citet{villamariz00}, we observe that the effect of $\varv_\text{turb}$ is only perceptible for hotter values of effective temperature (in our case, $T_\text{eff}\gtrsim33$ kK), primarily over the lines He I $\lambda$4387 and He I $\lambda$4922.
		\item The clumping factor $f_\text{cl}$ is fitted to the hydrogen lines, in particular, it fits the wings of H$\alpha$.
		As was pointed out by \citet{repolust04}, clumping effects over the wings of H$\alpha$ are more prominent when they are in emission than in absorption.
		In addition, we observe an effect over the core and the red wing over the line profiles of H$\beta$, H$\gamma$, H$\delta,$ and H$\epsilon$ as a function of the wavelength.
		Here, we follow the same notation as we used in Paper I that was taken from \citet{puls06}, where the clumping factor in FASTWIND is denoted as $f_\text{cl}$ and takes values $\ge1$ (where $f_\text{cl}=1$ is the value for the smooth limit).
		This $f_\text{cl}$ is then the inverse of the volume filling factor, denoted in CMFGEN as $f_\infty$ \citep{bouret12,alex21a}.
		This scale $f_\text{cl}=1/f_\infty$ relies on the assumption that the inter-clump medium is empty \citep{sundqvist18}.
		\item The macroturbulent velocity $\varv_\text{macro}$, which corresponds to the non-rotational contributions to the line broadening \citep[see][]{simon14}, is fitted to the width of He I lines, which have been empirically shown to be more sensitive to this broadening.
		\item The rotational velocity $\varv_\text{rot}$ is fitted to the slope in the shape of the wings of He II $\lambda$4200 and He II $\lambda$4541.
		For this purpose, the synthetic spectrum needs to be convolved after the end of the FASTWIND process.
		Moreover, we decided to make this convolution coherent with the effects of the $\Omega$ parameter over the self-consistent wind hydrodynamics,
		\begin{equation}
			\Omega=\frac{\varv_\text{rot}}{\varv_\text{crit}}\;,
		\end{equation}
		where $\varv_\text{crit}$ is the critical rotational speed for a spherical star \citep[see e.g.][for details]{araya17}.
		The rotational velocity is a third-order parameter, but $\Omega$ is a first-order parameter, which means that modifications in the convolution may require a complete restructuring of the self-consistent solution.
		Because we focus on the theoretical performance of synthetic spectra and because the focus is not on rapidly rotating stars, we prefer to denote rotational velocity simply as $\varv_\text{rot}$ instead of the classical $\varv\sin i$. 
		The convolution to broaden the spectral lines by rotational effects and the inclusion of the $\Omega$ factor over the hydrodynamic solution are both assumed to happen in the line of sight to the observer.
	\end{itemize}

	As a final remark, we mention that the radial velocity, $\varv_\text{rad}$, is obtained from the SIMBAD database\footnote{\url{http://simbad.u-strasbg.fr/simbad/}}.
	
%_____SPECTRAL RESULTS__________________________________________________________________________
\section{Spectral results}\label{spectralresults}

	In this section we present the results of the spectral analysis for the six stars of the sample, HD 192639, 9Sge, HD 57682, HD 218915, HD 195592, and HD 210809.
	For each of them, we present the best spectral fit (based on the trends introduced in Section~\ref{modificationparameters}) and a tabulation with the final stellar, wind, and line-force parameters found for these stars.
	An analysis and comparison of some of these new parameters with the literature, especially for the case of the mass-loss rate, are also considered.

	The spectra were fit with a by-eye inspection that searched for the model that globally satisfied all the criteria described above better.
	We selected this option instead of a $\chi^2$ reduction or an algorithm procedure \citep[e.g.  ][]{mokiem05} because an inspection by eye allows a more qualitative discussion of the fits.
	Of the 15 lines introduced in Table~\ref{fastwindlines}, the focus is mostly given to the fit of the wings for the set of hydrogen lines and to the emission peak of H$\alpha$.
	For helium, good-quality fits are expected for all the lines, except for He II $\lambda4686$ and He I $\lambda6678$, whose appropriate synthetic reproduction has been a challenge for the codes FASTWIND and CMFGEN \citep{puls05,holgado18}.
	The short execution time of the combined m-CAK procedure with FASTWIND (Fig.~\ref{mcakfastwind_flowchart}), which takes about one hour, allows developing multiple self-consistent wind solutions and then calculating the best model fit is quickly achieved with $\sim30$ model runs.

	Same as for Tables~\ref{tabletemperatures} and \ref{tablekk17}, the self-consistent wind parameters introduced in this section are presented with the error bars derived from the uncertainties of the line-force fitting and those derived from differences in the numbers of depth points used for the wind structure, as outlined in Section~\ref{errorbars}.
	Variance in the wind parameters due to uncertainties in the fitted stellar parameters ($T_\text{eff}$, $\log g$, He/H, etc.), which is a priori assumed to be about $\pm0.125$ for $\log\dot M$, is not included in these error bars.

%_____HD 192639__________________________________________________________________________________
\subsection{HD 192639}
	\begin{figure*}[t!]
		\centering
		\includegraphics[width=\linewidth]{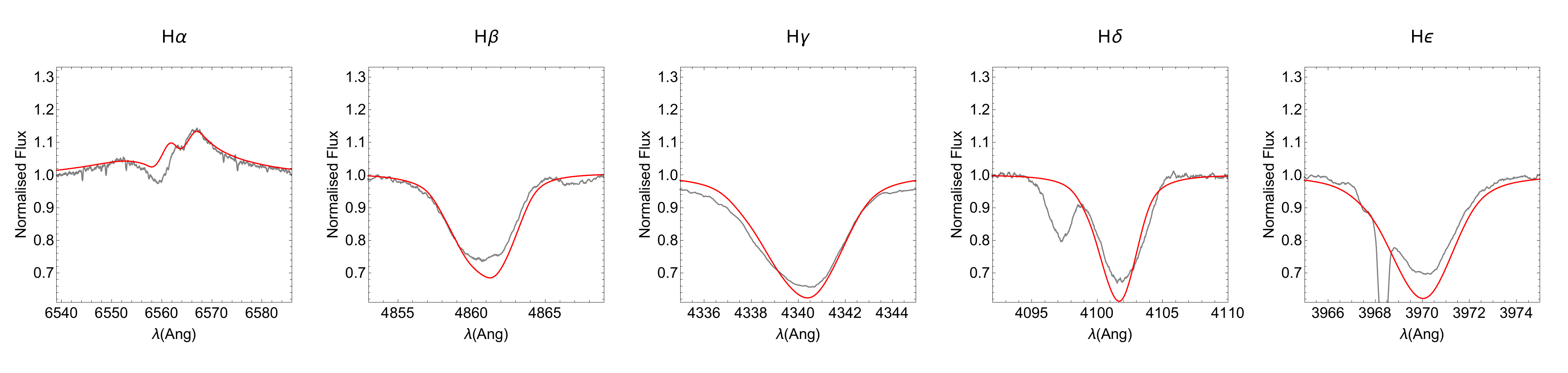}
		\includegraphics[width=\linewidth]{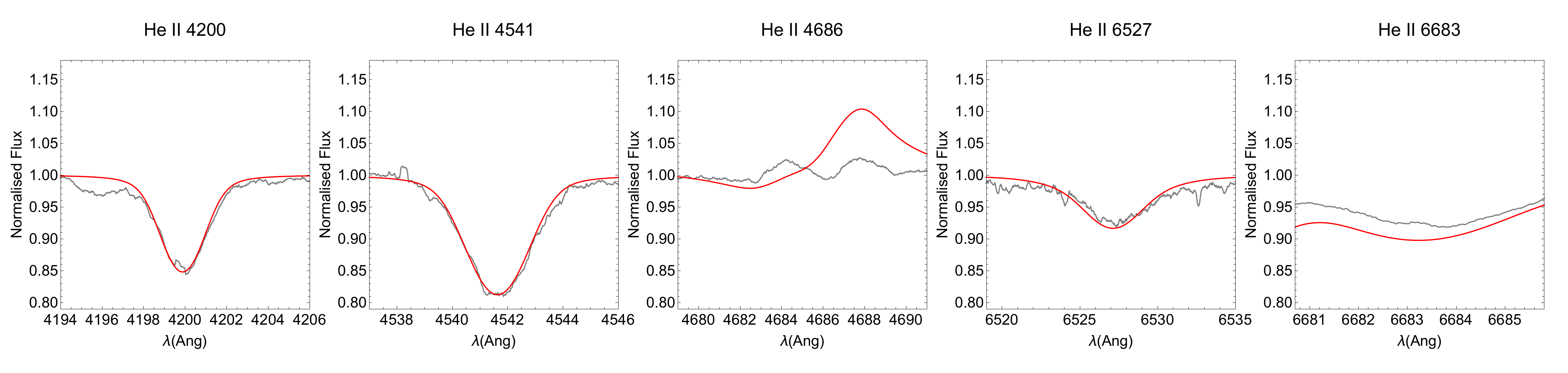}
		\includegraphics[width=\linewidth]{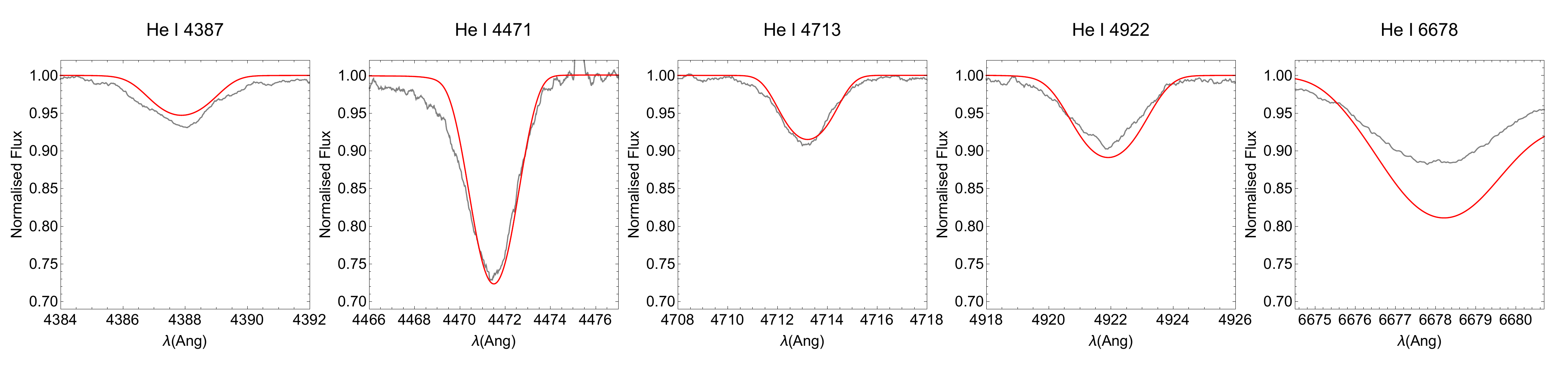}
		\caption{\small{Best fit for HD 192639 with the self-consistent solution tabulated in Table~\ref{tablehd192639}.}}
		\label{hd192639m042}
	\end{figure*}
	\begin{table}[h!]
		\centering
		\caption{\small{Summary of the stellar and wind parameters we used to fit HD 192639 (Fig.~\ref{hd192639m042}).}}
		\begin{tabular}{cc}
			\hline\hline
			\multicolumn{2}{c}{Parameters HD 192639}\\%hd192639m142
			\hline
			$T_\text{eff}$ (kK) & 34.0\\
			$\log g$ & 3.25\\
			$R_*/R_\odot$ & 19.8\\
			$M_*/M_\odot$ & $25.4$\\
			$L_*/L_\odot$ & $4.73\times10^5$\\
			$[\text{He/H}]$ & 0.10\\
			$(k,\alpha,\delta)_\text{sc}$ & $(0.047,0.694,0.089)$\\
			$\Omega$ & 0.26\\
			$\log\dot M$ ($M_\odot$ yr$^{-1}$) & $-5.783\pm.090$\\
			$\varv_\infty$ (km s$^{-1}$) & $1\,460\pm160$\\
			$f_\text{cl}$ & 6.25\\
			$\varv_\text{rad}$ (km s$^{-1}$) & 20\\
			$\varv_\text{rot}$ (km s$^{-1}$) & 100\\
			$\varv_\text{turb}$ (km s$^{-1}$) & 10\\
			$\varv_\text{macro}$ (km s$^{-1}$) & 30\\
			$\log D_\text{mom}$ & 28.83\\
			\hline
		\end{tabular}
		\label{tablehd192639}
	\end{table}

	The first star we analysed was the blue supergiant O7.5 I, HD 192639, observed with \textsc{Hermes} in 2013.
	This star was previously studied by \citet{bouret12}, who fitted a CMFGEN synthetic spectra with parameters $T_\text{eff}=33.5$ kK, $\log g=3.42$, $\log\dot M=-5.68,$ and $\beta=1.3$.
	We used these parameter as a starting point to perform our calculations instead of those derived from the calibrations by \citet{martins05} for a standard O 7.5 I star ($T_\text{eff}=32$ kK and $\log g=3.36$), which differ moderately, but still agree closely.
	The stellar parameters  we derived are tabulated in Table~\ref{tablehd192639} and also lie close to their previous values.
	The synthetic spectra are shown in Fig.~\ref{hd192639m042}.

	The initial rotational velocity for HD 192639 is slightly increased from $\varv_\text{rot}=90$ km s$^{-1}$ given by \citet{bouret12} to 100 km s$^{-1}$.
	This rotational velocity corresponds to an angular velocity of $\Omega=0.26$, which has a real impact on the generated velocity profile and on the line profiles, making them broader.
	This additional rotational broadening, which comes not only from the final convolution over the spectra, but also from the self-consistent solution, can partly replace the macroturbulent velocity, which is set to $\varv_\text{macro}=30$ km s$^{-1}$ instead of the 43 km s$^{-1}$ given by \citet{bouret12}.
	Another important upgrade: the self-consistent solution provides a higher effective temperature than the initial value, whereas the surface gravity is $\sim10\%$ lower.
	The final difference with the CMFGEN fit is our reduction of the He to H ratio from 0.15 to 0.1, which is necessary to reduce the intensity of all helium lines.

	The spectral fit we achieved shows a good match and is particularly good for the emission component of H$\alpha$, and with the remarkable exception of He II $\lambda\lambda$ 4686 and its inherent variability, the match is of the same order of accuracy as \citet[][see their figure~A.16]{bouret12}.
	Despite the mentioned variability, this fit gives us confidence for the self-consistent value we found for the mass-loss rate, $\log\dot M=-5.783$.
	This is just $\sim38\%$ higher ($+0.14$ dex in logarithmical scale) than the value found by \citet{bouret12}, but we used a clumping factor that is 3.2 times lower.

%_____9 Sge______________________________________________________________________________________
\subsection{9 Sge}
	\begin{figure*}[t!]
		\centering
		\includegraphics[width=\linewidth]{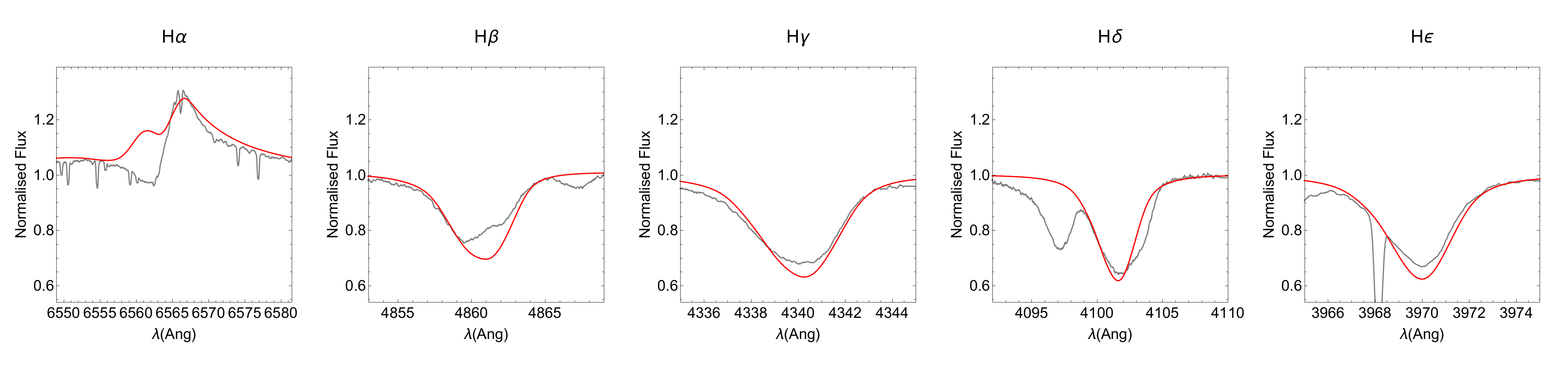}
		\includegraphics[width=\linewidth]{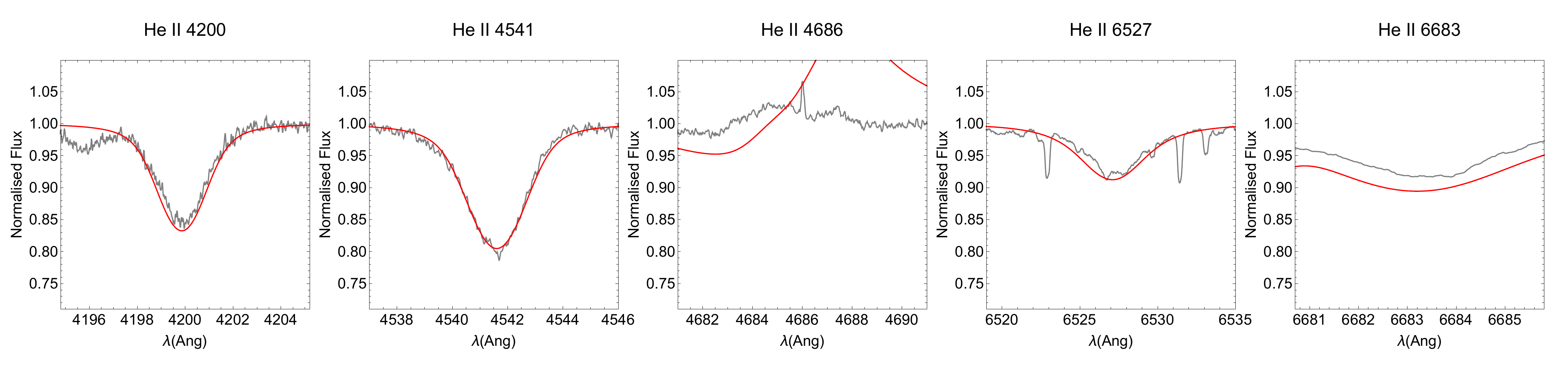}
		\includegraphics[width=\linewidth]{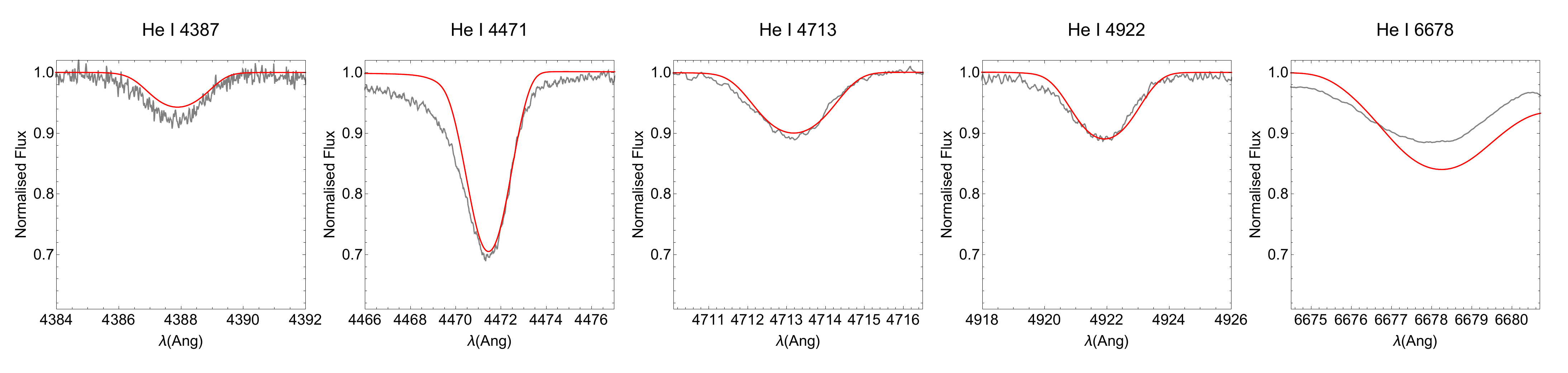}
		\caption{\small{Best fit for 9 Sge with the self-consistent solution tabulated in Table~\ref{table9sge}.}}
		\label{9sgem416}
	\end{figure*}
	\begin{table}[h!]
		\centering
		\caption{\small{Summary of the stellar and wind parameters we used to fit 9 Sge (Fig.~\ref{9sgem416}).}}
		\begin{tabular}{cc}
			\hline\hline
			\multicolumn{2}{c}{Parameters 9 Sge}\\%9sgem425
			\hline
			$T_\text{eff}$ (kK) & 34.5\\
			$\log g$ & 3.32\\
			$R_*/R_\odot$ & 23.0\\
			$M_*/M_\odot$ & $40.3$\\
			$L_*/L_\odot$ & $6.76\times10^5$\\
			$[\text{He/H}]$ & 0.125\\
			$(k,\alpha,\delta)_\text{sc}$ & $(0.057,0.685,0.075)$\\
			$\Omega$ & 0.21\\
			$\log\dot M_\text{sc}$ ($M_\odot$ yr$^{-1}$) & $-5.632\pm.073$\\
			$\varv_\infty$ (km s$^{-1}$) & $2\,000\pm190$\\
			$f_\text{cl}$ & 16.0\\
			$\varv_\text{rad}$ (km s$^{-1}$) & 43\\
			$\varv_\text{rot}$ (km s$^{-1}$) & 90\\
			$\varv_\text{turb}$ (km s$^{-1}$) & 20\\
			$\varv_\text{macro}$ (km s$^{-1}$) & 25\\
			$\log D_\text{mom}$ & 29.14\\
			\hline
		\end{tabular}
		\label{table9sge}
	\end{table}
	9 Sge (HD 188001) is a runaway star \citep{underhill95} with a spectral type of O 7.5 Iab \citep{sota11}.
	Do not confuse 9 Sge (from the Sagitta constellation) with 9 Sgr (from the Sagittarius constellation).
	This latter one is also an O starm but belongs to the main sequence spectral type O 3.5 V \citep{sota14}.
	9 Sge exhibits a periodic variation in its radial velocity of $P=97.6$ days, but it is still debated whether this period is produced by the presence of a companion \citep{maiz19}.
	Based on its spectral classification, the initial stellar parameters taken from the catalogue of \citet{martins05} are:$T_\text{eff}=34$, $\log g=3.36$ and $R_*=20.8\,R_\odot$, which are also the parameters provided by the VizieR Online Data Catalog\footnote{\url{http://vizier.u-strasbg.fr/viz-bin/VizieR?-source=J/A+A/620/A89}} \citep{nebot18}.
	Alternatively, \citet{martins15a} found $T_\text{eff}=33$, $\log g=3.35$.
	The fitted parameters and the spectrum are presented in Table~\ref{table9sge} and Fig.~\ref{9sgem416} respectively.

	To achieve a closer agreement for the helium lines, it was necessary to increase the He to H ratio from the solar standard from \citet[][He/H=0.085]{asplund09} up to He/H=0.125.
	Moreover, wide shape of the line profiles led us to require an increase in both macro-turbulence (25 km s$^{-1}$) and rotational velocity (90 km s$^{-1}$), together with perform wind solutions with a lower gravity.
	Same as for the case of HD 192639, H$\alpha$ is fitted by its wings and by the emission component, whereas the absorption is only partially reproduced.
	The self-consistent mass-loss rate obtained is $2.33\times10^{-6}$ $M_\odot$ yr$^{-1}$, roughly $\sim1.8$ times higher than valued provide by \citet{nebot18}.
	
%_____HD 57682__________________________________________________________________________________
\subsection{HD 57682}
	\begin{figure*}[t!]
		\centering
		\includegraphics[width=\linewidth]{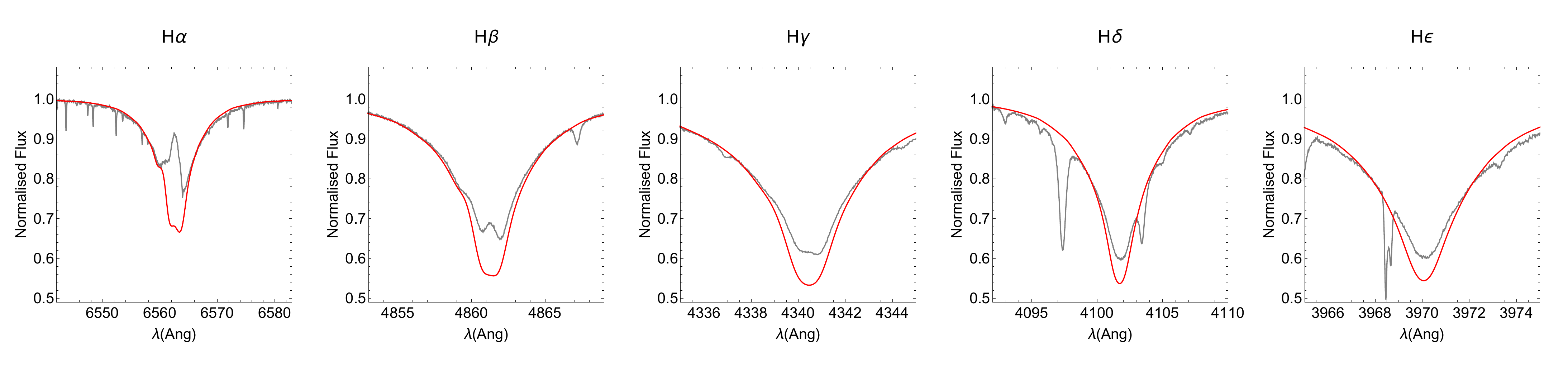}
		\includegraphics[width=\linewidth]{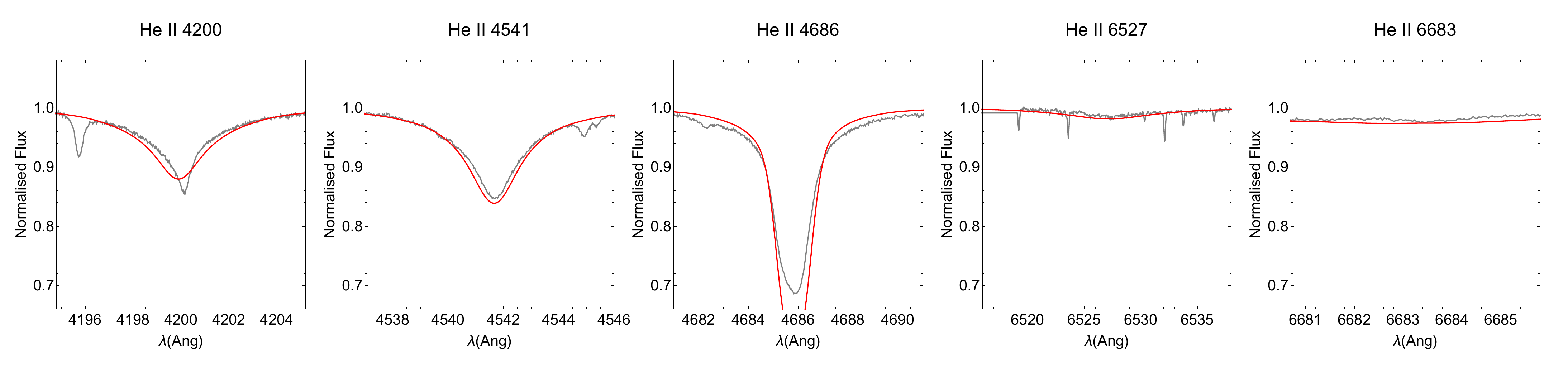}
		\includegraphics[width=\linewidth]{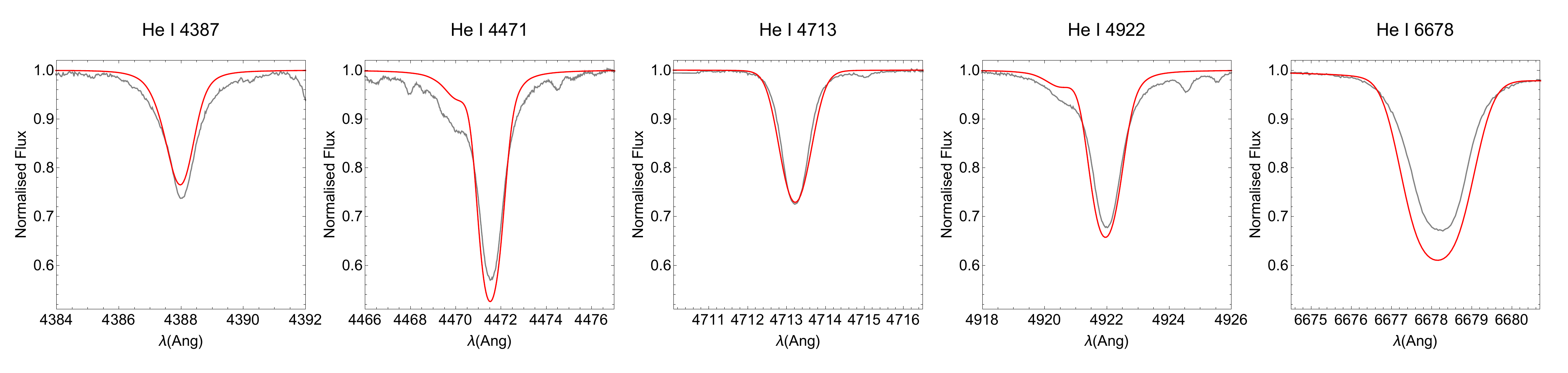}
		\caption{\small{Best fit for HD 57682 with the self-consistent solution tabulated in Table~\ref{tablehd57682}.}}
		\label{hd57682m037}
	\end{figure*}
	\begin{table}[h!]
		\centering
		\caption{\small{Summary of the stellar and wind parameters we used to fit HD 57682 (Fig.~\ref{hd57682m037}).}}
		\begin{tabular}{cc}
			\hline\hline
			\multicolumn{2}{c}{Parameters HD 57682}\\
			\hline
			$T_\text{eff}$ (kK) & 35.5\\
			$\log g$ & 3.85\\
			$R_*/R_\odot$ & 7.0\\
			$M_*/M_\odot$ & $12.7$\\%\pm2.6
			$L_*/L_\odot$ & $7.02\times10^4$\\%\pm2.6
			$[\text{He/H}]$ & 0.085\\
			$\Omega$ & 0.02\\
			$(k,\alpha,\delta)_\text{sc}$ & $(0.117,0.629,0.092)$\\
			$\log\dot M_\text{sc}$ ($M_\odot$ yr$^{-1}$) & $-6.759\pm.085$\\
			$\varv_{\text{sc},\infty}$ (km s$^{-1}$) & $2\,020\pm120$\\
			$f_\text{cl}$ & 2.5\\
			$\varv_\text{rad}$ (km s$^{-1}$) & 20\\
			$\varv_\text{rot}$ (km s$^{-1}$) & 15\\
			$\varv_\text{turb}$ (km s$^{-1}$) & 20\\
			$\varv_\text{macro}$ (km s$^{-1}$) & 20\\
			$\log D_\text{mom}$ & 27.75\\
			\hline
		\end{tabular}
		\label{tablehd57682}
	\end{table}

	We modelled the wind of the O9 star HD 57682, which is the only non O-supergiant star of our \textsc{Hermes} sample.
	It was originally classified as luminosity-class IV by \citet{walborn72}, but reclassified as main sequence (luminosity-class V) by \citet{grunhut12}.
	The study of \citet{grunhut09} found line profile variability (LPV) due to a magnetic field $B\sim1600\,G$.
	Despite this high value for $B$, these studies have successfully achieved accurate spectral fittings using the spherically symmetric code CMFGEN.
	Based on this study, we started with the following stellar parameters: $T_\text{eff}=35$ kK, $\log g=4.0,$ and $R_*=7\,R_\odot$.
	In this particular case, the self-consistent solution obtained from this set of stellar parameters quickly leads to a good fit after little tuning of the effective temperature and surface gravity.
	The spectrum is presented in Fig.~\ref{hd57682m037}, and the obtained parameters are shown in Table~\ref{tablehd57682}.

	\citet{grunhut12} found that the most prominent LPV for HD 57682 is observed for H$\alpha$, where a strong emission peak in the core of the line arises.
	Moreover, the circumstellar material surrounding the star also contaminates the core of the other hydrogen lines such as H$\gamma$ \citep[][see their Fig.2]{grunhut09}.
	For the case of helium lines, we derive a slight increase in the effective temperature of ($500$ K) and an also slight decrease in surface gravity ($0.1$ dex), which helps to improve the fit for He I lines in particular (He I $\lambda4471$, probably because of its variable behaviour).

	For the mass-loss rate, \citet{grunhut09} determined a value of $\log\dot M=-8.85$ from the C IV lines $\lambda1548$ and $\lambda1551,$ but they stated that the mass-loss rate derived from the emission component of H$\alpha$ should be $\sim10^{-7}$ solar masses per year.
	The peculiar emission profiles for H$\alpha$ were later reproduced in some detail by \citet[][see their Fig.~15]{grunhut12}, based on the magnetohydrodynamic (MHD) simulations performed by \citet{sundqvist12}, where the mass-loss rate was readapted to $\log\dot M\sim-7.73$.
	This means that the emission component of HD 57682 is only reproduced in 2D models.
	It is important to note that these values from previous authors were all calculated assuming a $\beta$-law for the wind velocity profile using the code CMFGEN \citep{hillier90b,hillier98}.
	In contrast,  the theoretical mass-loss rate ($\log\dot M_\text{sc}=-6.759$) we derived is able to fit the wings of H$\alpha$ down to $\sim\pm100$ km s$^{-1}$ along the line core, even though it is $\sim120$ times higher than the initially established
value.
	This shows the difference between velocity profiles that are calculated hydrodynamically with radiative acceleration and $\beta$-law profiles, although it would be necessary to perform self-consistent MHD calculations to study the full shape of H$\alpha$ in magnetic stars such as HD 57682 in more detail in the future.

	Finally, we also mention the issue of the macroturbulence.
	For their CMFGEN models, \citet{grunhut09} adopted a value of $\varv_\text{macro}=40$ km s$^{-1}$, whereas in \citet{grunhut17} the authors adopted an even higher value of 65 km s$^{-1}$.
	For our spectra, we fit the wings of He I lines using a lower value for the macroturbulence of $\varv_\text{macro}=20$ km s$^{-1}$, which does not dispose any overestimation of the rotational velocity \citep[we found $\varv_\text{rot}=15$ km s$^{-1}$, which is of the same magnitude order as the 13 km s$^{-1}$ from][]{grunhut12}.

%_____HD 218915_________________________________________________________________________________
\subsection{HD 218915}\label{hd218915}
	\begin{figure*}[t!]
		\centering
		\includegraphics[width=\linewidth]{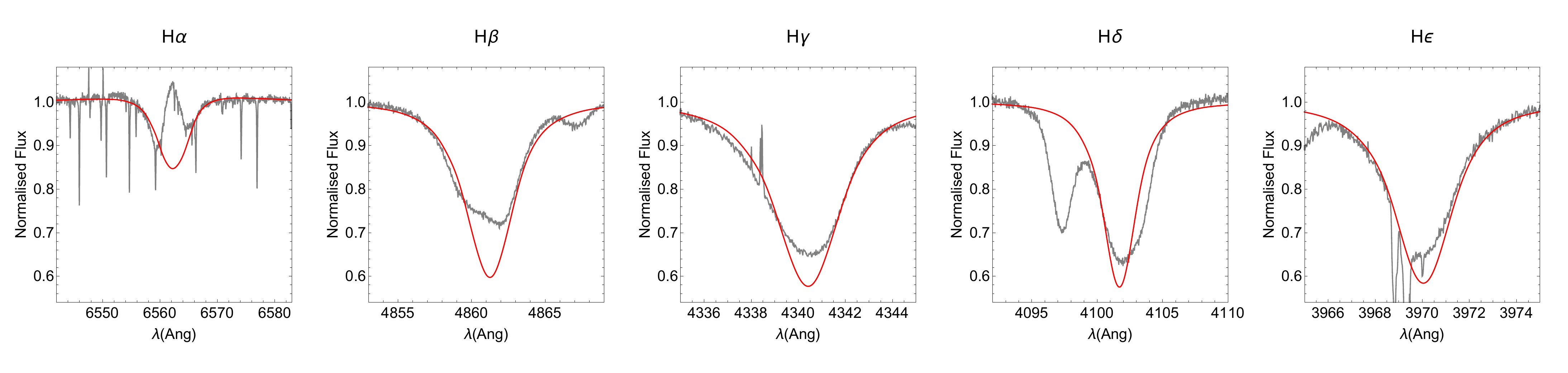}
		\includegraphics[width=\linewidth]{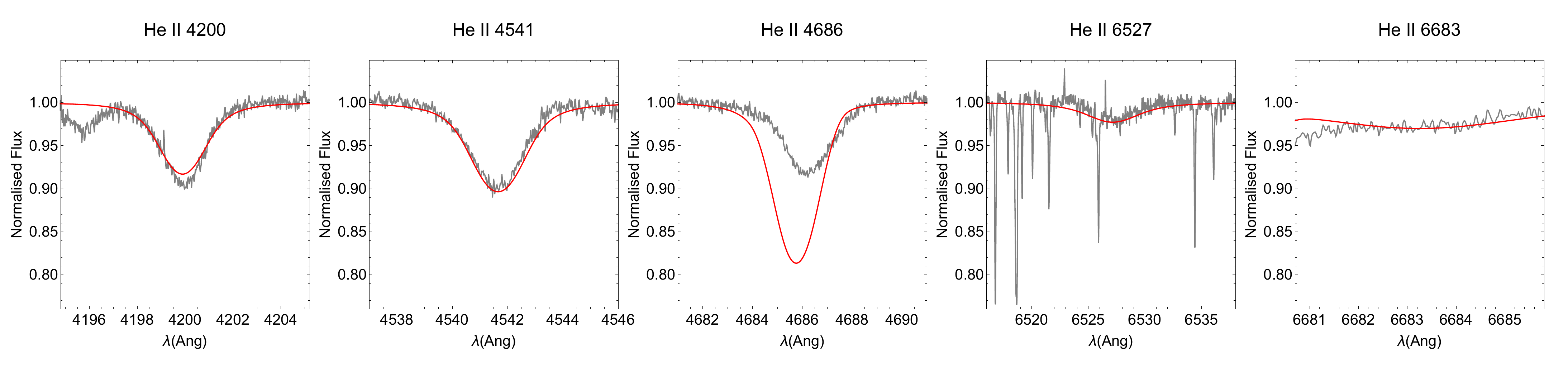}
		\includegraphics[width=\linewidth]{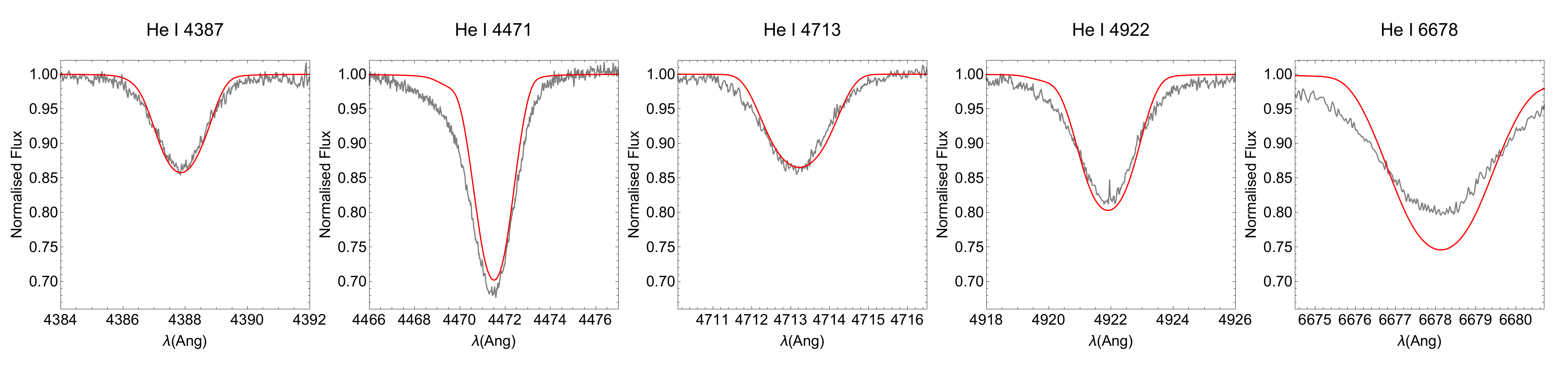}
		\caption{\small{Best fit for HD 218915 with the self-consistent solution tabulated in Table~\ref{tablehd218915}.}}
		\label{hd218915m633}
	\end{figure*}
	\begin{table}[h!]
		\centering
		\caption{\small{Summary of the stellar and wind parameters we used to fit HD 218915 (Fig.~\ref{hd218915m633}).}}
		\begin{tabular}{cc}
			\hline\hline
			\multicolumn{2}{c}{Parameters HD 218915}\\%hd218915m633
			\hline
			$T_\text{eff}$ (kK) & 31.0\\
			$\log g$ & 3.23\\
			$R_*/R_\odot$ & 18.0\\
			$M_*/M_\odot$ & $20.1$\\%\pm2.6
			$L_*/L_\odot$ & $2.7\times10^5$\\%\pm2.6
			$[\text{He/H}]$ & 0.07\\
			$\Omega$ & 0.2\\
			$(k,\alpha,\delta)_\text{sc}$ & $(0.057,0.690,0.124)$\\
			$\log\dot M_\text{sc}$ ($M_\odot$ yr$^{-1}$) & $-6.015\pm.106$\\
			$\varv_{\text{sc},\infty}$ (km s$^{-1}$) & $1\,390\pm200$\\
			$f_\text{cl}$ & 1.67\\
			$\varv_\text{rad}$ (km s$^{-1}$) & $-95$\\
			$\varv_\text{rot}$ (km s$^{-1}$) & 80\\
			$\varv_\text{turb}$ (km s$^{-1}$) & 20\\
			$\varv_\text{macro}$ (km s$^{-1}$) & 15\\
			$\log D_\text{mom}$ & 28.55\\
			\hline
		\end{tabular}
		\label{tablehd218915}
	\end{table}
	
	HD 218915 is a blue supergiant that was classified as O 9.5 Iab by \citet{petit19}, and as O 9.2 Iab by \citet{martins18}.
	Like 9 Sge, HD 218915 is a runaway star with a periodicity in its radial velocity of $P=0.89$ days \citep{barannikov06}.
	From \citet{leitherer88}, we can obtain the following parameters to set our initial calculations: $M_*=45\,M_\odot$, $R_*=26\,R_\odot$ and $\log L_*/L_\odot=5.62$, whereas the study performed by \citet{holgado18} provides $T_\text{eff}=31.1$ kK and $\log g=3.21$.
	The final parameters we calculated for the self-consistent solution are tabulated in Table~\ref{tablehd218915}, and the spectral fit is shown in Fig.~\ref{hd218915m633}

	The most prominent aspect to be noted from the new parameters we derived from our self-consistent procedure is the apparent reduction of the He to H ratio to 0.07: this is $\sim18\%$ below the new ratio established by solar metallicity by \citet{asplund09}.
	This is a remarkable situation because it is commonly assumed that Galactic stars are born with solar metallicity, and the abundance of helium in their surface and wind will only be enhanced through their evolutionary tracks due to nucleosynthesis \citep[][in prep.]{ekstrom12,alex22}, implying therefore that HD 218915 might have been born with a peculiarly low He/H ratio.
	Nevertheless, this modification was absolutely necessary to fit the faint absorption profiles of the helium lines, as the decrement on surface was saturating the lines.
	This was confirmed after we ran several models to scan the entire parameter space (including $\varv_\text{turb}$).
	An alternative explanation for the depleted helium of HD 218915 is binarity: the line might become diluted by the light of a potential companion.
	However, because this is a runaway star, this hypothesis does not seem to apply in this particular case.
	The implications of this behaviour deserve a particular study, which is beyond the scope of this work.
	Regarding other stellar parameters, both $T_\text{eff}$ and $\log g$ lie below the threshold rigidly stated in Paper I (same as the supergiant 325-I from Table~\ref{tablekk17}), but even so, an adequate spectral fit for the star was achieved, therefore we can establish self-consistent wind solutions under m-CAK prescription remain a valid framework for effective temperatures and surface gravities below 32 kK and 3.4.
	
	Similarly as HD 57682, HD 218915 exhibits an emission component in H$\alpha$.
	This might be an indication for the existence of circumstellar material surrounding the star, again due to magnetic activity.
	However, studies from \citet{grunhut17} and \citet{petit19} did not find evidence of a strong magnetic field for HD 218915, so that the origin of this additional emission component in H$\alpha$ remains unclear.
	This additional component has not been reproduced before, even using a $\beta$-law \citep{holgado18}.
	Therefore, its origin can be attributed to the variability of the star.
	Although it has been possible to accurately reproduce an H$\alpha$ emission component in another O9.5 I star using synthetic spectra \citep[see e.g. the case of HD 188209 in][]{martins15b}, this was achieved using an extremely high value of $\beta=2.2$ for the velocity profile \citep[far from the $\beta=0.8-1.2$ used by][]{holgado18}.
	In the case of HD 188209, the rotational velocity was also lower ($\varv\sin i=45$ km s$^{-1}$).
	Because H$\beta$ and H$\gamma$ also show a weak additional component at the core, the wing profiles of these Balmer lines were fitted.
	The self-consistent mass-loss rate, $\log\dot M_\text{sc}=-6.015$ agrees surprisingly well with the value provided by \citet{leitherer88}.
	This might partly be attributed to the reduction of the stellar radius made by us from 26 down to 18 $R_\odot$, which is crucial to reduce the intensity of He I lines.
	The broadness of He I profiles also was satisfied by increasing the rotational velocity up to 80 km s$^{-1}$, in contrast with the $~60$ km s$^{-1}$ reported by previous studies \citep{leitherer88,grunhut17,petit19} and especially assuming a macroturbulence of 15 km s$^{-1}$, far below the 94 km s$^{-1}$ reported by \citet{grunhut17}.
	
%_____HD 195592_________________________________________________________________________________
\subsection{HD 195592}\label{hd195592}
\begin{figure*}[t!]
		\centering
		\includegraphics[width=\linewidth]{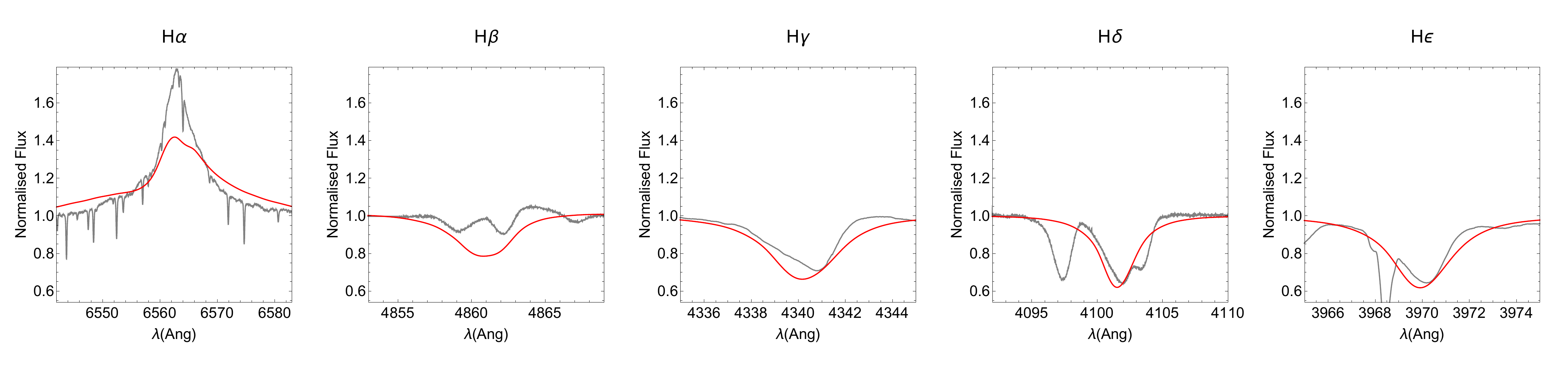}
		\includegraphics[width=\linewidth]{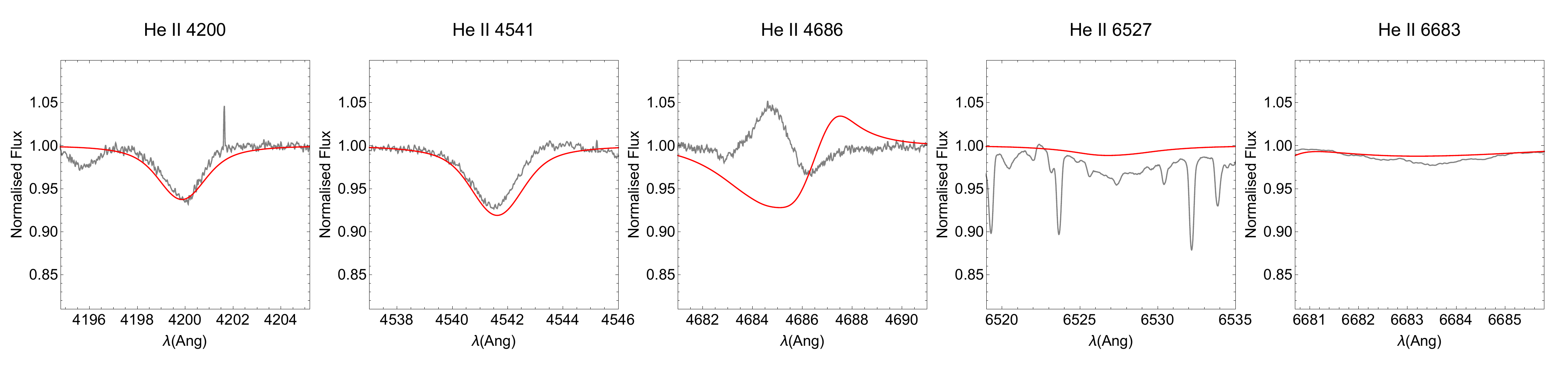}
		\includegraphics[width=\linewidth]{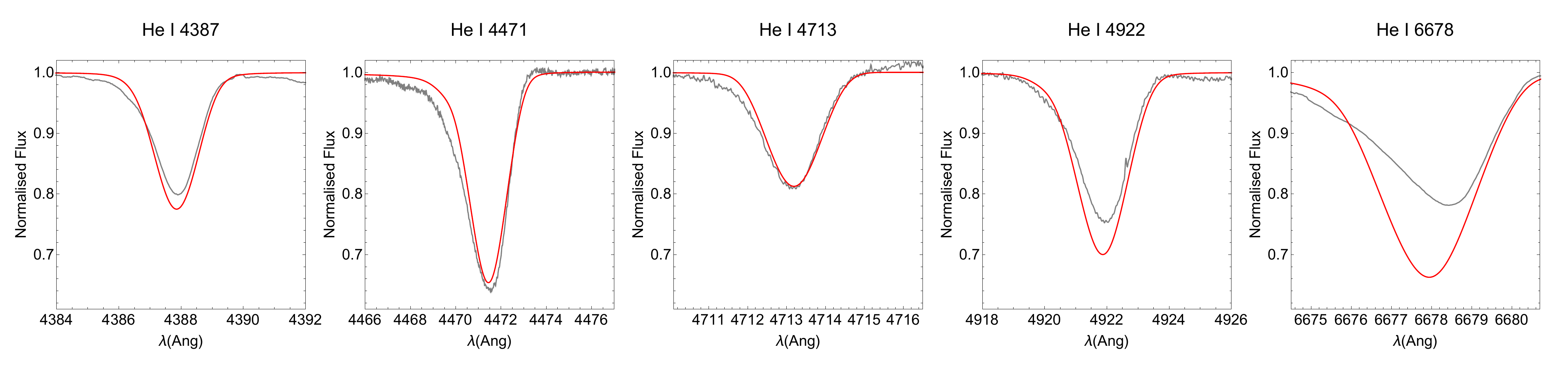}
		\caption{\small{Best fit for HD 195592 with the self-consistent solution tabulated in Table~\ref{tablehd195592}.}}
		\label{hd195592m322}
	\end{figure*}
	\begin{table}[h!]
		\centering
		\caption{\small{Summary of the stellar and wind parameters we used to fit HD 218915 (Fig.~\ref{hd195592m322}).}}
		\begin{tabular}{cc}
			\hline\hline
			\multicolumn{2}{c}{Parameters HD 195592}\\%hd195592m322
			\hline
			$T_\text{eff}$ (kK) & 29.5\\
			$\log g$ & 3.20\\
			$R_*/R_\odot$ & 21.5\\
			$M_*/M_\odot$ & $26.7$\\%\pm2.6
			$L_*/L_\odot$ & $3.16\times10^5$\\
			$[\text{He/H}]$ & 0.095\\
			$\Omega$ & 0.15\\
			$(k,\alpha,\delta)_\text{sc}$ & $(0.128,0.670,0.180)$\\
			$\log\dot M_\text{sc}$ ($M_\odot$ yr$^{-1}$) & $-5.369\pm.557$\\
			$\varv_{\text{sc},\infty}$ (km s$^{-1}$) & $1\,390\pm120$\\
			$f_\text{cl}$ & 1.0\\
			$\varv_\text{rad}$ (km s$^{-1}$) & $-5$\\
			$\varv_\text{rot}$ (km s$^{-1}$) & 60\\
			$\varv_\text{turb}$ (km s$^{-1}$) & 10\\
			$\varv_\text{macro}$ (km s$^{-1}$) & 25\\
			$\log D_\text{mom}$ & 29.18\\
			\hline
		\end{tabular}
		\label{tablehd195592}
	\end{table}
	
	HD 195592 is another runaway star, classified as O 9.7 Ia \citep{sota11}.
	It presents evidence of being a binary system, which can explain the line profile variations (LPV) observed for the spectrum \citep{debecker10}.
	The initial mass $M_*=30\,M_\odot$, radius $R_*=30\,M_\odot$ , and luminosity $L_*=3.1\times10^5$ $L_\odot$ were taken from \citet{debecker10}, who derived them from \citet{martins05} given its spectral classification for the closest subtype to HD 195592 (O 9.5 I).
	We derived our initial parameters also from that calibration, while the final values are tabulated in Table~\ref{tablehd195592} and the synthetic spectra are shown in Fig.~\ref{hd195592m322}.
	
	The search for a self-consistent solution for this star led us to find the most extreme parameters of our sample: $T_\text{eff}=29.5$ kK and $\log g=3.2$.
	To achieve this result, it was necessary to increase the number of points for our stratificated hydrodynamic solution to 999 to avoid crashes and false convergences.
	We call a false convergence the finding of different sets of $(k,\alpha,\delta)$ given different initial values $(k_0,\alpha_0,\delta_0)$.
	This situation violates the basic principle of uniqueness for a self-consistent solution: \textit{one and only one} set of line-force parameters satisfies the e.o.m and recovers the same line-acceleration \citep[see Fig.~2 from][]{alex19}.
	As we showed in Fig.~\ref{npoints_wind}, the selection of the number of points produces significant variations in the resulting wind solution when effective temperature and surface gravity take too low values.
	Although this discrepancy was observed more over the terminal velocity than over the mass-loss rate, it is evident that the resulting hydrodynamics have an impact on the future synthetic spectra to be generated, which can create a false positive depending on the topology of the solution \citep{michel07}.
	A deeper analysis is required for a better understanding of the reasons for this exotic behaviour, but this is beyond the scope of this study (see the discussion in Section~\ref{multipleselfconsistency}).
	The importance of this situation is that the applicability of the m-CAK self-consistent wind prescription is less confident for effective temperatures below 30 kK, and therefore this value could be reconsidered as a new threshold for the recipe.
	This hypothesis is reinforced by the fit obtained for H$\alpha$, where we can infer that the self-consistent velocity profile for a wind solution with a cool temperature of $T_\text{eff}=29.5$ kK is not adequate to shape the wings and core of the line. \citet{marcolino17} provided an even lower effective temperature ($T_\text{eff}=28$ kK) and a lower surface gravity ($\log g=2.9$), which confirms that HD 195592 lies outside the range where m-CAK self-consistent solutions can currently be found.
	In any case, an emission in H$\alpha$ is at least predicted, and the mass-loss rate provided by $\dot M\sqrt{f_\text{cl}}$ is relatively close to the homogeneous value provided by \citet[][$\log\dot M_\text{hom}=-5.14$]{marcolino17}.
	
%_____HD 210809________________________________________________________________________________
\subsection{HD 210809}\label{hd210809}
	\begin{figure*}[t!]
		\centering
		\includegraphics[width=\linewidth]{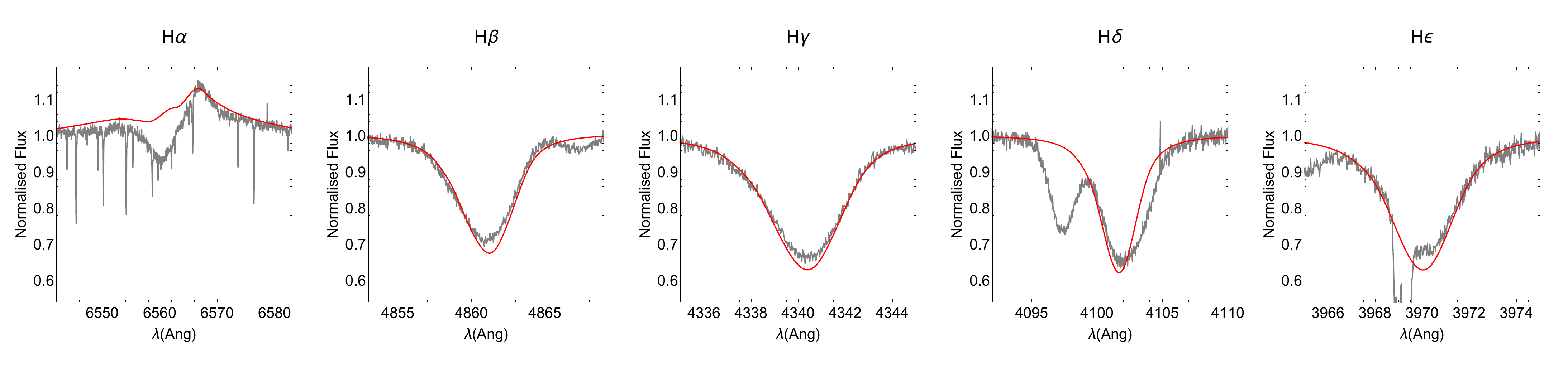}
		\includegraphics[width=\linewidth]{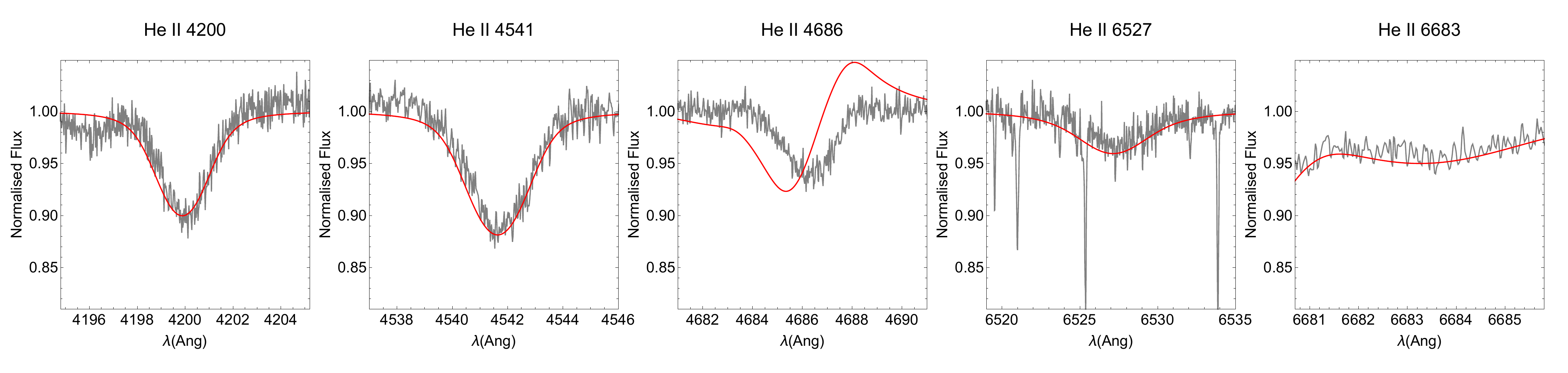}
		\includegraphics[width=\linewidth]{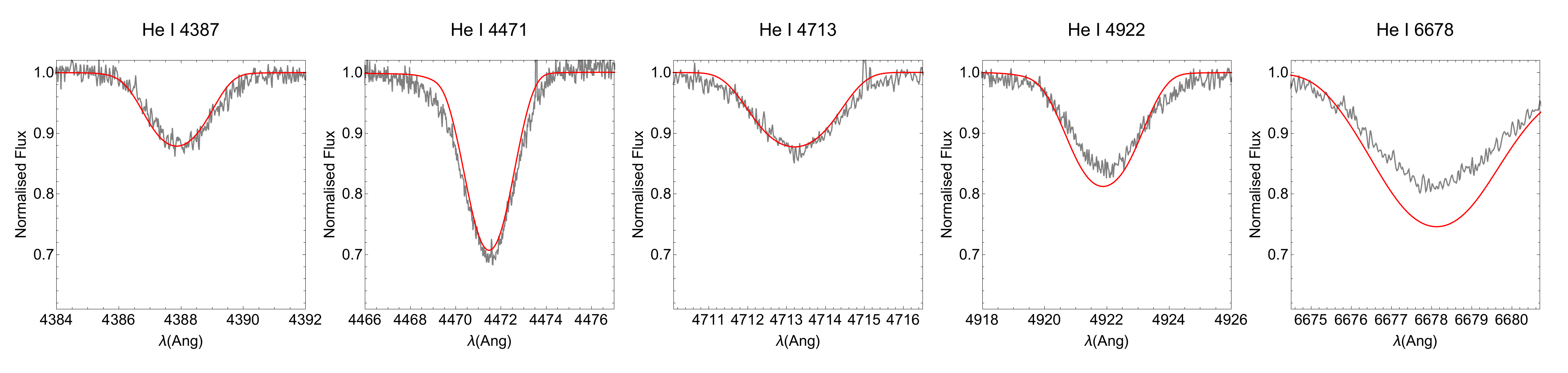}
		\caption{\small{Best fit for HD 210809 with the self-consistent solution tabulated in Table~\ref{tablehd210809}.}}
		\label{hd210809m522}
	\end{figure*}
	\begin{table}[h!]
		\centering
		\caption{\small{Summary of the stellar and wind parameters we used to fit HD 210809 (Fig.~\ref{hd210809m522}).}}
		\begin{tabular}{cc}
			\hline\hline
			\multicolumn{2}{c}{Parameters HD 210809}\\%hd210809m522
			\hline
			$T_\text{eff}$ (kK) & 31.5\\
			$\log g$ & 3.20\\
			$R_*/R_\odot$ & 22.0\\
			$M_*/M_\odot$ & $28.0$\\%\pm2.6
			$L_*/L_\odot$ & $4.3\times10^5$\\
			$[\text{He/H}]$ & 0.10\\
			$\Omega$ & 0.2\\
			$(k,\alpha,\delta)_\text{sc}$ & $(0.068,0.688,0.141)$\\
			$\log\dot M_\text{sc}$ ($M_\odot$ yr$^{-1}$) & $-5.593\pm.207$\\
			$\varv_{\text{sc},\infty}$ (km s$^{-1}$) & $1\,340\pm120$\\
			$f_\text{cl}$ & 2.2\\
			$\varv_\text{rad}$ (km s$^{-1}$) & $-70$\\
			$\varv_\text{rot}$ (km s$^{-1}$) & 100\\
			$\varv_\text{turb}$ (km s$^{-1}$) & 20\\
			$\varv_\text{macro}$ (km s$^{-1}$) & 25\\
			$\log D_\text{mom}$ & 28.98\\
			\hline
		\end{tabular}
		\label{tablehd210809}
	\end{table}
	
	The last star of our sample is HD 210809, another blue supergiant, classified as O 9 Ia \citep{sota11}.
	According to this, the stellar parameters in the catalog of \citet{martins05} are $31.3$ kK for the effective temperature, $3.23$ for $\log g,$ and $21.76$ for the stellar radius, but the survey of \citet{petit19} provided a value of $R_*=15.2\,R_\odot$.
	Final parameters calculated for the self-consistent solution are tabulated in Table~\ref{tablehd210809}, whereas spectral fit is shown in Fig.~\ref{hd210809m522}.
	
	The rotational velocity was assumed to be 100 km s$^{-1}$ \citep{grunhut17}.
	This makes HD 210809, together with HD 192639, the fastest rotators of our sample.
	This is also clear in the P-Cygni profile observed for the H$\alpha$ line in both stars.
	Hence the fit of H$\alpha$ is focused on the emission component, as for the HD 192639 case (Fig.~\ref{hd192639m042}).
	Although we do not reproduce the absorption component of H$\alpha$, the accurate fit achieved for the rest of the Balmer series makes us confident about the solution found for HD 210809.
	
	The self-consistent value for mass-loss rate found for HD 210809, $\dot M=2.55\times10^{-6}$ $M_\odot$ yr$^{-1}$, is $\sim30\%$ lower than the value of $\log\dot M=-5.44$ from \citet{fullerton06}, but it agrees with the theoretical value from the usage of the Vink formulae with the stellar parameters tabulated in Table~\ref{tablehd210809} ($\dot M_\text{Vink}=2.5\times10^{-6}$ $M_\odot$ yr$^{-1}$).
	On the other hand, KK17 provided a theoretical value of $\dot M_\text{KK17}=1.9\times10^{-7}$ $M_\odot$ yr$^{-1}$.
	Therefore, as shown in Table~\ref{tablekk17}, the m-CAK self-consistent mass-loss closely agrees with Vink's formula for the cases of $T_\text{eff}\sim32$ kK.

%_____DISCUSSION_________________________________________________________________________________
\section{Discussion}\label{discussion}

%_____Multiple self-consistent solution and saturation
\subsection{Multiple self-consistent solution and saturation}\label{multipleselfconsistency}
	\begin{table*}[t!]
		\centering
		\caption{\small{Line-force parameters calculated for the O 9.7 I star HD 195592 (see Section~\ref{hd195592}), given different numbers of depth points and different sets of effective temperature and surface gravity.}}
		\begin{tabular}{ccccc}
			\hline\hline
			\multicolumn{5}{c}{HD 195592 (false solutions)}\\
			& & \multicolumn{3}{c}{$(k,\alpha,\delta)_\text{sc}$}\\
			$T_\text{eff}$ & $\log g$ & \multicolumn{3}{c}{Nº depth points}\\
			$[\text{kK}]$ & & $200$ & $500$ & $999$\\
			\hline
			29.5 & 3.2 & $(0.059,0.695,0.127)$ & $(0.065,0.683,0.135)$ & $(0.128,0.670,0.180)$\\
			& & $(0.078,0.692,0.147)$\\
			& & $(0.108,0.690,0.170)$\\
			\hline
			30.0 & 3.2 & $(0.060,0.684,0.127)$ & $(0.096,0.682,0.155)$ & $(0.135,0.683,0.190)$\\
			& & $(0.118,0.685,0.186)$\\
			\hline
			30.5 & 3.2 & $(0.144,0.704,0.195)$ & $(0.145,0.703,0.194)$ & $(0.149,0.701,0.193)$\\
			\hline
			29.5 & 3.22 & $(0.048,0.685,0.106)$ & $(0.054,0.684,0.116)$ & $(0.119,0.667,0.173)$\\
			\hline
		\end{tabular}
		\label{falsehd195592}
	\end{table*}

	As we pointed out in Fig.~\ref{npoints_wind}, the self-consistent solutions for line-force parameters depend on the number of depth points, which produce variations in the resulting wind parameters to be calculated.
	These variations look more profound for the case of low temperatures, where our standard model of $T_\text{eff}=32$ kK and $\log g=3.4$ showed large uncertainties in the resulting self-consistent terminal velocity.
	
	This erratic behaviour is also expressed in the uncertainty for the self-consistent mass-loss rates (our most important wind parameter) when the effective temperature and the surface gravity decreases the threshold imposed in Paper I.
	These uncertainties are expressed in the error bars associated with the derived self-consistent $\dot M$, due to a larger contribution of the term $\Delta_{N_\text{points}}^2$ in Eq.~\ref{deltawind}, which are now in the order of the errors assumed to be produced by uncertainties in the stellar parameters such as in the case of HD 218915 (Section~\ref{hd218915}, see also the considerations incorporated in Section~\ref{errorbars}).
	
	However, as we previously pointed out for the analysis of the star HD 195592 (Section~\ref{hd195592}), a decrement in the temperature not only produced a broadening in the error bar for the mass-loss rate, but also a problem in finding a unique solution for any number of depth points implemented.
	These are the so-called false convergences, and they are better illustrated in Table~\ref{falsehd195592}, where we present different sets of self-consistent line-force parameters $(k,\alpha,\delta)$ given different numbers of depth points.
	When we use $T_\text{eff}\le30$ kK and $\log g=3.2,$ we observe more than one solution for the case of the lower number of points (200). This situation does not appear with a larger grid for the wind structure.
	Even though false solutions disappear when the number of depth points is increased, however, we still observe a huge discrepancy between the sets of $(k,\alpha,\delta)$ calculated, which makes the error bar in these cases $\sim0.5$ (such as for HD 195592).
	A more detailed analysis of the coupling of the line-acceleration and the topology of hydrodynamics is required to understand the phenomenon that is generated for temperatures $<30$ kK, but this is beyond the scope of this work.
	At this point, we can state that the discrepancies in the resulting self-consistent wind parameters cause us to set our threshold of the validity for our m-CAK self-consistent prescription at $T_\text{eff}=30$ kK and $\log g=3.2$.
	
	Another important case to consider is HD 210809 (Section~\ref{hd210809}).
	The effective temperature is above our threshold of 30 kK, but we still found huge divergences in the self-consistent solutions when we modified the number of depth points.
	Particularly for HD 210809, we find that these large differences appear when the rotation velocity reaches the value $\Omega=0.2$.
	This is better displayed in Table~\ref{saturatedhd210809}, where below this value, variations in the line-force parameters lie within the normal discrepancies due to the selection of different number of depth points; but above this limit, every set of self-consistent $k$, $\alpha,$ and $\delta$ changes dramatically.
	Moreover, for $\Omega>0.2$, increasing the number of the depth points always produces the same self-consistent solution where the $\delta$ parameter reaches the value of $0.204$, which makes it closer to the so-called `delta-slow solutions' \citep{michel11,venero16,araya21}.
	We call this a `saturated' solution because it is independent of the initial value for $\Omega>0.2$ and because the resulting wind solution (with a mass-loss rate of $\dot M\simeq1.4\times10^{-5}$ $M_\odot$ yr$^{-1}$) cannot generate a synthetic spectrum.
	Therefore, analogously to the previous case of HD 195592, we found a region in the set of initial stellar parameters in which the self-consistent m-CAK prescription presents problems to provide a unique well-converged solution, this time related with a high rotational velocity.
	However, given that HD 210809 is not our fastest rotator of the sample (the fastest is HD 192639 with $\Omega=0.26$, see Table~\ref{tablehd192639}), we guess that its low effective temperature ($T_\text{eff}=31.5$) might act as another threshold for the cases when the rotational velocity becomes significant.
	In a forthcoming paper, we expect to explore the effects of the rotational velocity on the resulting self-consistent solutions with effective temperatures below 32 kK in more detail.
	Hence, we consider that for stars with $\Omega>0.2$, the range of validity of our m-CAK prescription is restricted to $T_\text{eff}\gtrsim31.5$ kK.

	\begin{table}[t!]
		\centering
		\caption{\small{Line-force parameters calculated for the O 9 Ia star HD 210809 (see Section~\ref{hd210809}), given different numbers of depth points and different sets of effective temperature and surface gravity.}}
		\begin{tabular}{cccc}
			\hline\hline
			\multicolumn{4}{c}{HD 210809 (saturated solutions)}\\
			& \multicolumn{2}{c}{$(k,\alpha,\delta)_\text{sc}$} &\\
			& \multicolumn{2}{c}{Nº depth points}\\
			$\Omega$ & $200$ & $300$\\
			\hline
			0.25 & $(0.068,0.688,0.141)$ & $(0.126,0.716,0.204)$\\
			0.23 & $(0.068,0.689,0.142)$ & $(0.126,0.716,0.204)$\\
			0.21 & $(0.066,0.689,0.140)$ & $(0.125,0.716,0.204)$\\
			0.20 & $(0.065,0.689,0.137)$ & $(0.070,0.687,0.143)$\\
			0.0 & $(0.055,0.693,0.115)$ & $(0.057,0.689,0.118)$\\
			\hline
		\end{tabular}
		\label{saturatedhd210809}
	\end{table}
	
	Nevertheless, despite the false convergence and saturation discussed in this section, we rely on the final parameters introduced for HD 195592 and HD 210809 in Tables~\ref{tablehd195592} and \ref{tablehd210809}, respectively.
	They represent the best global spectral fit we achieved for each star, as shown in Figures~\ref{hd195592m322} and \ref{hd210809m522}.

%_____Limits of the self-consistent m-CAK prescription and the issue of clumping
\subsection{Limits of the self-consistent m-CAK prescription and the issue of clumping}
	The work introduced in Paper I has provided a grid of wind solutions to self-consistently calculate the line-force parameters for the line-acceleration with hydrodynamics.
	Because line-force parameters come from the m-CAK framework, these wind solutions are calculated following the formal assumptions of the theory such as the Sobolev approximation, and neglect the detailed influence of multi-scattering effects.
	Despite these issues, the m-CAK prescription from Paper I has been shown to provide values for $\dot M$ in agreement with homogeneous mass-loss rates derived from observations \citep{bouret05,markova18}, in addition to the calculation of reliable synthetic spectra to fit clumped stellar winds \citep[see section 6.2 from][]{alex19}.
	Now, the inclusion of a stratified temperature for the stellar wind, based on the prescription used by \citet{sundqvist19} and \citet{lucy71}, has led to a decrement in the self-consistent values for the mass-loss rate, particularly for effective temperatures of $T_\text{eff}\sim40$ kK or higher, mostly because of the decrement in the line-force parameter $k$.
	Decrements in $\dot M,$ however, still agree with theoretical values presented by other self-consistent studies such as KK17 \citep{kk17}, as shown in Table~\ref{tablekk17}.
	On the one hand, this would be expectable if we were to consider that both studies used flux fields from \textsc{Tlusty} \citep{hubeny95}.
	On the other hand, the fact that the results are close, even considering that these authors calculated their line-acceleration from CMF radiative transfer equations (whereas we calculate the line-force parameters), may suggest that some of the drawbacks commonly attributed to the m-CAK framework \citep[such as line-blanketing and multi-scattering effects, see][]{puls87} are not relevant when self-consistent calculations are implemented (at least in the range of temperatures we study), as was previously argued in \citet[][Section~2]{alex19}.
	
	Moreover, regardless of the differences with Paper I, the implementation of the temperature structure $T(r)$ shows successful spectral fittings with our sample of stars, as was demonstrated in Section~\ref{spectralresults}.
	This could be discussed in favour of the correctness of the self-consistent values for $\dot M$ presented in this work, but in this case, we would also require a proper theoretical determination of the clumping factor for each model.
	It has been shown that the implementation of different clumping factors leads to different line-accelerations, which in turn will lead to a different self-consistent solution for the wind \citep{alex21a}.
	Including these effects on the calculated $g_\text{line}$ from the line-force parameters would require a meticulous description of the inhomogeneities through the wind structure, such as the 3D Monte-Carlo calculations developed by \citet{surlan12,surlan13}.
	However, these calculations also use a $\beta$-law prescription for their velocity profiles, so that a new iterative procedure would be required to couple the two calculations and then to obtain a self-consistent clumping description and $\varv(r)$.
	We expect to move in this direction in a forthcoming study, but this is currently beyond of the scope of this work.
	
	Even when a fully consistent calculation of every parameter of the stellar wind is the most desirable scenario and the final goal of every theoretical prescription, in practice, this achievement cannot be reached.
	The spectral fits we presented and their respective self-consistent wind parameters $\dot M_\text{sc}$ and $\varv_{\infty,\text{sc}}$ were made on the basis that the fitted stellar parameters are correct, especially because the error bars associated with each wind parameter do not consider the uncertainties of temperature, mass, radius, or abundances.
	This is the price to pay for reducing the number of free parameters for the spectral fit.
	Therefore, we do not claim that our predicted mass-loss rate and terminal velocity are `the truth', but we state that our prescription leads to reliable theoretical wind parameters that can adequately reproduce (despite particular exceptions such as the H$\alpha$ profile of HD 218915) spectral observations and agree with other studies.
	Based on these results, self-consistent m-CAK wind solutions show a promising perspective for the theoretical determination of mass-loss rates, and hence we are confident in the potential of this prescription in the future.

\subsection{Accuracy of spectral fitting}
	\begin{figure*}[t!]
		\centering
		\includegraphics[width=0.45\linewidth]{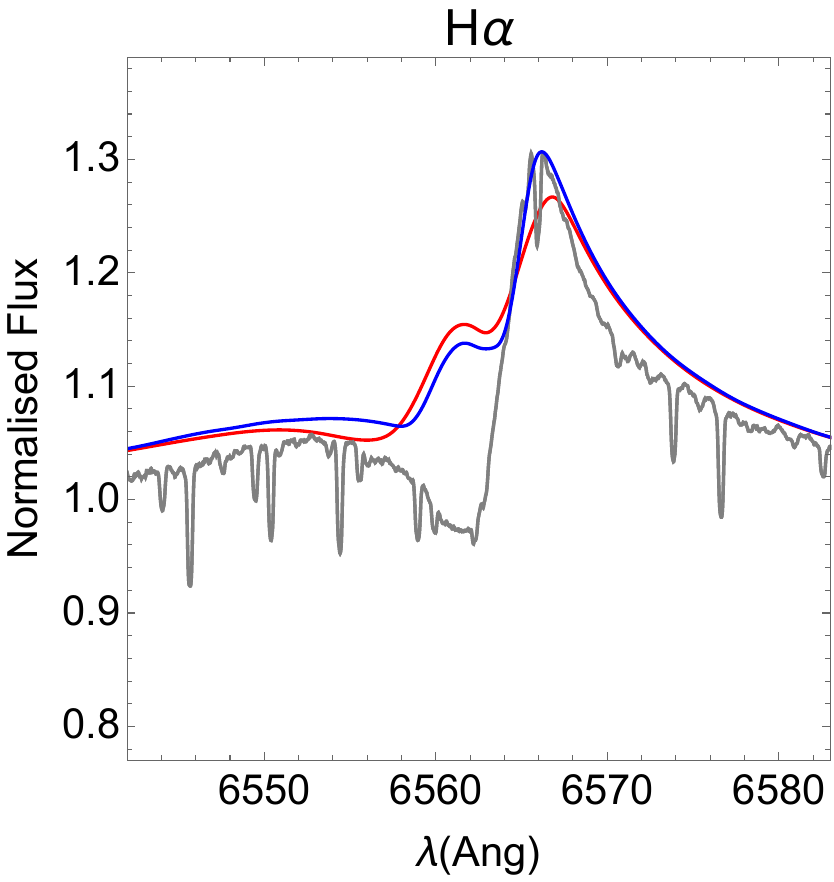}
		\hspace{1cm}
		\includegraphics[width=0.45\linewidth]{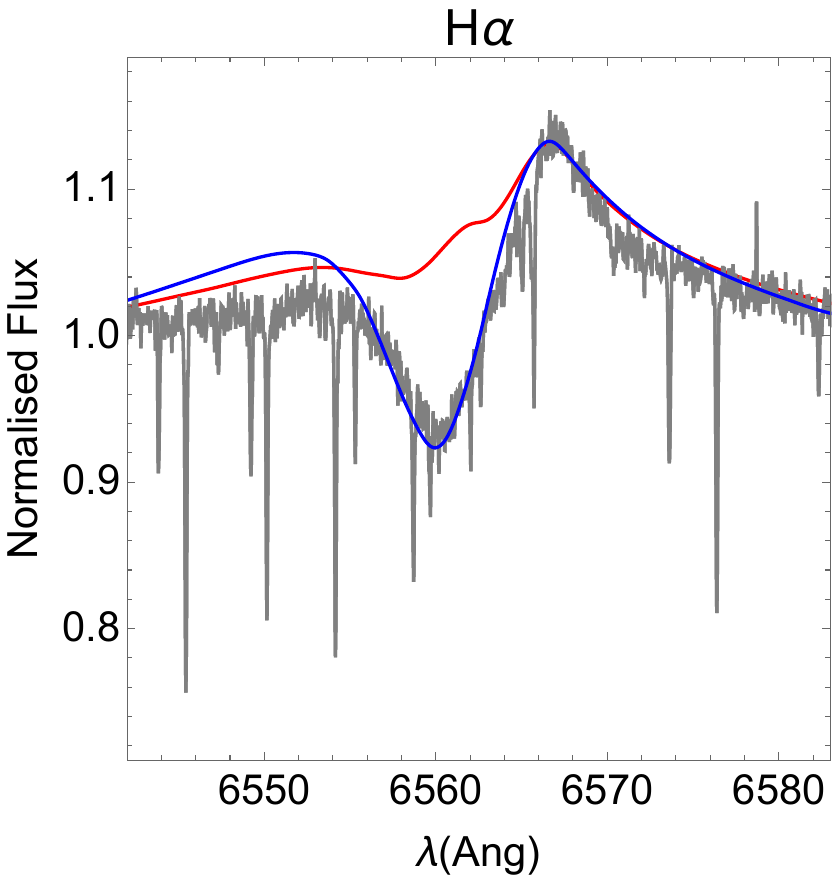}
		\caption{\small{Comparison of the best fits for 9 Sge (left-panel) and HD 210809 (right-panel), with the self-consistent solution introduced in Fig.~\ref{9sgem416} for 9 Sge in Fig.~\ref{hd210809m522} for HD 210809 (red) and with the improved fits individualised for H$\alpha$ (blue).}}
		\label{fitHalphas}
	\end{figure*}

	In Section \ref{spectralresults} we introduced a set of stellar and wind parameters to simultaneously fit a sample of optical spectra observed with the \textsc{Hermes} spectrograph.
	The fit was achieved by eye, and our main criterion was the global accuracy for the lines tabulated in Table~\ref{fastwindlines}.
	However, despite our general focus on providing accurate fits for most of the listed lines (mainly the photospheric lines), the quality the fits achieved for some of the H$\alpha$ profiles in our sample might not be considered satisfactory.
	
	Improved fits for the H$\alpha$ lines of 9 Sge and HD 210809 are shown in Fig.~\ref{fitHalphas}.
	For the case of 9 Sge, we can improve the fit of H$\alpha$ if we increment the turbulence velocity to 40 km s$^{-1}$ and reduce the rotational velocity to 50 km s$^{-1}$.
	In addition, the macroturbulence needs to be decreased to 4 km s$^{-1}$ in order to properly reproduce the sharpness of the H$\alpha$ profile.
	This improvement produces a sharper emission component, even when it was not possible to reproduce the absorption portion of the P-Cygni profile of the line.
	For the case of HD 210809, a more accurate fit is achieved also by setting $\varv_\text{turb}=40$ km s$^{-1}$, decreasing $\varv_\text{macro}=2$ km s$^{-1}$ and increasing the rotational velocity to $\varv_\text{rot}=170$ km s$^{-1}$.
	In this case, the absorption and emission components of the line are accurately reproduced, but these new modifications do not fit the other spectral lines.
	
	Therefore, the fits introduced in Section~\ref{spectralresults} are the best global fits provided by our method when we aim to obtain a simultaneous fit for the visible range of the spectra rather than focus only on one single line such as H$\alpha$.
	Studies focused on the self-consistent description of the wind either do not provide a spectral fit \citep{kk17,bjorklund21} or the presented spectral fits do not allow a flexible fine-tuning \citep{sander17}.
	The reason for this dichotomy is, as stated in \citet{alex21a}, that a perfect fully self-consistent wind prescription must consider every feature present in the wind (e.g. the clumping prescription and even effects from X-rays), which is beyond the scope of this work.
	In the case of our m-CAK prescription, even when the selection of the wind parameters is based on a unique solution coming from stellar parameters (and then they are not free parameters), the subsequent spectral fit is capable to adequately reproduce the lines in the optical range.
	Some lines cannot be adjusted, such as the H$\alpha$ emission components for HD 57682 and HD 218915, because spherically symmetric codes were used, but this is not a failure attributable to the m-CAK method. As shown by \citet{grunhut12}, CMFGEN is also not able to fit these additional emission components.
	The exception is the poor fit for H$\alpha$ in HD 195592, but as discussed in Section~\ref{hd195592}, this is an indicator that the limits of our prescription coincide with the lower threshold of $T_\text{eff}\simeq30$ kK, as stated in Section~\ref{temperaturerange}.
	Because of these reasons, we are confident that our spectral fitting using the m-CAK prescription is reliable, and therefore the stellar and wind parameters determined by our method are trustworthy.
	We expect in the future to extend this analysis to UV spectra, where P-Cygni profiles such as O V $\lambda$1371, C IV $\lambda1548,$ and N IV $\lambda$1718 will be useful to evaluate the terminal wind velocities and clumping factors \citep{bouret12,surlan12}.

\subsection{Self-consistent WML relation}
	The wind-momentum-luminosity (WML) relation predicts a strong dependence of the wind momentum rate on the stellar luminosity with an exponent determined by the statistics of the strengths of the hundreds of thousands of lines driving the wind \citep[see][]{puls00}.
	It also contains a weak dependence on stellar radius.
	This is due to the work against the gravitational potential that the stellar wind must overcome. 
	This relation has been confirmed observationally using objects from the Milky Way, the LMC, SMC, and other galaxies of the local group by \citet{bresolin01,bresolin03}, \citet{urbaneja02,urbaneja03}, \citet{evans05,evans07}, and \citet{mokiem05,mokiem06}.
	
	According to \citet{kudritzki00}, $D_\text{mom}$ is defined as
	\begin{equation}
		D_\text{mom}=\dot M\varv_\infty\sqrt{R_*/R_\odot}\;,
	\end{equation}
	and the WML reads
	\begin{equation}
		\log D_\text{mom}=\log D_0+x\,\log(L_*/L_\odot)\;.
		\label{eqWML}
	\end{equation}
	
\begin{figure}[htbp]
		\centering
		\includegraphics[width=0.9\linewidth]{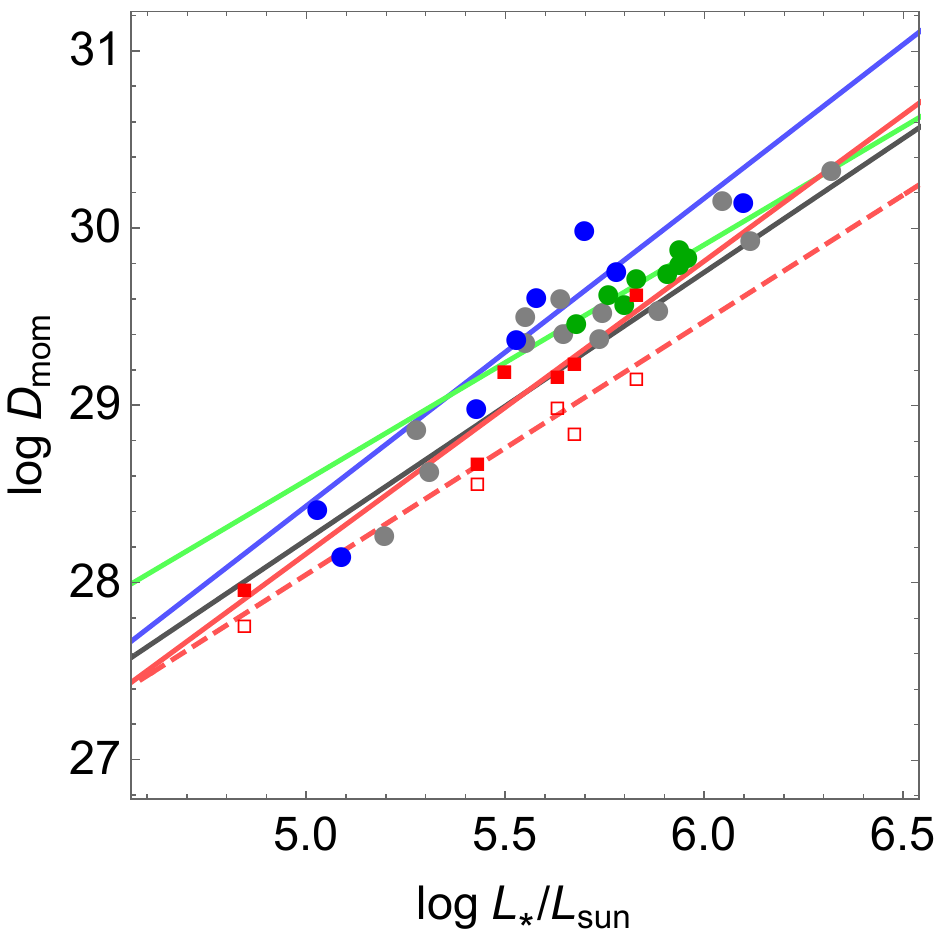}
		\caption{\small{Wind momemtum, $D_\text{mom}$ for the stars of our sample as a function of their luminosity. Filled red squares represent the momentum derived from a homogeneous mass-loss rate (see Eq.~\ref{dmomhom}), and empty red squares represent the clumped values for $\dot M$. In addition, we overplot the results for O I stars from \citet[][grey circles, from their Fig. 8]{kudritzki00}, the Galactic O-supergiants from \citet[][green circles]{bouret12}, and the field stars taken from \citet[][blue circles]{markova18}. The respective linear regressions from Table~\ref{dmomtable}, following the same colour convention, are also included.}}
		\label{dmom_sample}
	\end{figure}

	Figure \ref{dmom_sample} shows the WML for the sample of O I stars from \citet{kudritzki00} as grey circles and our results as red squares.
	Additionally, we included $D_\text{mom}$ values for the sample of field stars from \citet{bouret12}, whose values for mass-loss rate and terminal velocities are based on spectral fitting; and \citet{markova18}, which are based on unclumped values for the mass-loss rate and hence were previously used to compare with self-consistent solutions \citep[][Fig.~13]{alex19}.
	Although our sample is reduced, our $\log D_\text{mom}$ clearly accurately matches the values from \citet{kudritzki00}.
	To make this comparison coherent, we used unclumped values for our self-consistent mass-loss rates following
	\begin{equation}\label{dmomhom}
		D_\text{mom,hom}=\dot M_\text{clump}\,f_\text{cl}^{1/2}\varv_\infty\sqrt{R_*/R_\odot}\;.
	\end{equation}
	
	Finally, Table \ref{dmomtable} compares the values of $\log D_0$ and $x$ (Eq. \ref{eqWML}) from \citet{kudritzki00} and from \citet{markova18} with our results.
	We observe that both \citet{bouret12} and \citet{markova18} list a deviation from the original fit of \citet{kudritzki00}, but are still within the limits of the standard deviations of $\log D_0$ and $x$.
	On the other hand, our homogeneous values for $D_\text{mom}$   follow almost the same trend as in the previous studies.
	This is particularly clear in Fig.~\ref{dmom_sample}, where empty squares (clumped $\dot M$) lie slightly out of the main band, but filled squares (unclumped $\dot M$) lie inside it.
	Despite these encouraging results, the fits introduced in Table~\ref{dmomtable} represent a simple linear regression based on a few points, and we certainly need to expand our sample and extend our analysis to UV wavelengths to provide a more accurate conclusion.
	
	\begin{table}[t!]
		\centering
		\caption{\small{Comparison of the coefficients of the wind momentum–luminosity.}}
		\resizebox{0.49\textwidth}{!}{
		\begin{tabular}{ccc}
			\hline\hline
			Source & $\log D_0$ & $x$\\
			\hline
			\citet[][O I]{kudritzki00} & $20.69\pm1.04$ & $1.51\pm0.18$\\
			\citet[][O III, V]{kudritzki00} & $19.87\pm1.21$ & $1.57\pm0.21$\\
			\citet{bouret12} (unclumped) & $21.93\pm1.08$ & $1.33\pm0.18$\\
			\citet{markova18} & $18.46\pm1.45$ & $1.96\pm0.26$\\
			This work (unclumped) & $19.91\pm1.16$ & $1.65\pm0.20$\\
			This work (clumped) & $20.91\pm1.74$ & $1.42\pm0.32$\\
			\hline
		\end{tabular}}
		\label{dmomtable}
	\end{table}

%_____CONCLUSIONS_______________________________________________________________________________
\section{Conclusions}\label{conclusions}
	
	We have exploited the potential of the m-CAK prescription performed by \citet[][Paper I]{alex19} to derive self-consistent wind parameters by performing spectral fittings for a sample of six O-type stars (HD 192639, 9Sge, HD 57682, HD 218915, HD 195592, and HD 210809) observed with the high-resolution echelle spectrograph \textsc{Hermes}.
	In addition, the m-CAK prescription has been improved to include a temperature structure throughout the wind, together with a more exhaustive quantitative implementation of the uncertainties associated with each model of self-consistent solution.
	
	In a broader context, self-consistent synthetic spectra agree well with \textsc{Hermes} observations, which are no longer based on a $\beta$-law for their velocity profiles.
	Improvements could be made using this method to fit H$\alpha$ in particular, but this would undermine the quality of the fit for the remaining lines.
	This success reinforces the validity of the set of new stellar and wind parameters introduced in this work for our sample of stars, especially when there are significant differences with previous literature.
	These differences can be attributed to the previous usage of a $\beta$-law, in contrast with our self-consistent hydrodynamics calculations.
	Nevertheless, our method treats clumping as a free parameter that is only modified when the line fits are fine-tuned (Section~\ref{modificationparameters}), even when it is known that clumps affect the line-acceleration \citep{alex21a} because inhomogeneities are currently not included when the line-force parameters are calculated (Section~\ref{mcakprescription}).
	A full implementation of clumping would require a thorough analysis to describe the wind inhomogeneities  \citep{surlan12,surlan13}.
	In addition, the inclusion of UV spectral analysis will complement the constraint of mass-loss rate and clumping.
	However, this is beyond the scope of this work, and here we just introduce the first attempt to derive stellar and wind parameters for O-type stars through a spectral analysis considering the self-consistent wind hydrodynamics.
	We expect that the values introduced here are helpful for future studies of all the stars constituting this sample.
	
	The qualitative diagnostics and the detailed analysis of the self-consistent wind solutions obtained for some stars establish that the validity of our m-CAK prescription applies to O-type stars with $T_\text{eff}\ge30$ kK and $\log g\ge3.2$ (although the threshold for effective temperature should increase if we consider rotational velocity $\Omega>0.2$).
	We remark that this range of validity is where the line-force parameters $(k,\alpha,\delta)_\text{sc}$ can be fitted from the force multiplier $\mathcal M(t)$ assuming them to be constants instead as a function of radius, because their variance is below the common uncertainties associated with stellar parameters.
	Subsequently, these thresholds indicate that self-consistent solutions can describe the winds of massive stars from their birth on the main sequence and cover then a large part of their lifetime in this stage before their evolution into later-type stars. This aspect is incorporated in \citet[][in preparation]{alex22}.

%_____AGRADECIMIENTOS___________________________________________________________________________
\begin{acknowledgements}
	We sincerely thank the anonymous referee for the valuable feedback and comments provided to us.
	We also thank J. Puls for helpful discussions that improved this work and for having put at our disposal his code FASTWIND.
	This project has received funding from the European Unions Framework Programme for Research and Innovation Horizon 2020 (2014-2020) under the Marie Sk{\l}odowska-Curie grant Agreement N$^{\rm o}$ 823734.
	ACGM has been financially supported by the PhD Scholarship folio N$^{\rm o}$ 21161426 from National Commission for Scientific and Technological Research of Chile (CONICYT).
	ACGM and MC acknowledge support from Centro de Astrof\'isica de Valpara\'iso.
	MC, IA and CA thank the support from FONDECYT project 1190485.
	AL acknowledges in part funding from a ESA/PRODEX Belgian Federal Science Policy Office (BELSPO) contract related to the Gaia Data Processing and Analysis Consortium.
	JC thanks the supports from FONDECYT project 1211429.
	JAP has been funded by grants from Universidad Nacional de La Plata, Argentina (Proyecto I+D 11/G162).
	IA thanks the support from FONDECYT project 11190147.
	CA thanks to FONDECYT project N. 11190945.
\end{acknowledgements}

%_____BIBLIOGRAFÍA_______________________________________________________________________________
\bibliography{hermespaper.bib} % your references Yourfile.bib
\bibliographystyle{aa} % style aa.bst

\begin{appendix} %First appendix
\section{Tau}\label{tau}
	The relation between the radiative acceleration and 
	\begin{equation}
		g_\text{rad}(r)=\frac{\kappa_\text{F}L_*}{4\pi cr^2}=\frac{\kappa_\text{F} F(r)}{c}\;,
	\end{equation}
	with
	\begin{equation}
		\kappa_\text{F}F(r)=\int_0^\infty\kappa_\nu F_\nu d\nu\;.
	\end{equation}
	
	\begin{eqnarray}
		\tau_\text{F}(r) & = & \int\kappa_\text{F}\,\rho(r)\left(\frac{R_*}{r}\right)^2\,dr\;,\nonumber\\
		& = & \int\frac{cg_\text{rad}}{F(r)}\rho(r)\left(\frac{R_*}{r}\right)^2\,dr\;,\nonumber\\
		& = & \int g_\text{rad}\frac{4\pi cr^2}{L_*}\rho(r)\left(\frac{R_*}{r}\right)^2\,dr\;,\nonumber\\
		& = & \frac{4\pi cR_*^2}{L_*}\int g_\text{rad}\,\rho(r)\,dr\;,\nonumber\\
		& = & \frac{4\pi cR_*^2}{L_*}\int (g_\text{line}+g_\text{es})\,\rho(r)\,dr\;,\nonumber\\
		& = & \frac{4\pi cR_*^2}{L_*}\int (g_\text{line}+g_\text{grav}\Gamma_\text{es})\,\rho(r)\,dr\;.
	\end{eqnarray}
	
	Therefore, for the temperature field,
	\begin{eqnarray}
		T(r) & = & T_\text{eff}\left[W(r)+\frac{3}{4}\tau_\text{F}\right]^{1/4}\;,\nonumber\\
		& = & T_\text{eff}\left[W+\frac{3\pi cR_*^2}{L_*}\int (g_\text{line}+g_\text{grav}\Gamma_\text{es})\,\rho\,dr\right]^{1/4}\;.
	\end{eqnarray}
	
	To numerically solve the integral, see Appendix~\ref{numericalintegral}.

\section{Numerical integral}\label{numericalintegral}
	We have the integral
	\begin{equation}
		\int_r^\infty (g_\text{line}+g_\text{grav}\Gamma_\text{es})\,\rho(r')\,dr'\;.
	\end{equation}
	
	Being $x(r)=g_\text{rad}\,\rho(r)$ and $X=\int xdr$, the integral can be rewritten in a discrete way as
	\begin{equation}
		X_i=\sum_i^N\frac{x_i+x_{i+1}}{2}\times(r_{i+1}-r_i)\;.
	\end{equation} 
	The problem with this formulation is that for the last item of the grid (i.e. when $i=N$),  $\Delta r$ has no definite value because the limit is the infinite.
	In this case, it is better to perform a change of variables over the integral.
	Using the auxiliar variable $u=-1/r$ and therefore
	\begin{equation}
		\int_{-1/u}^0 (g_\text{line}+g_\text{grav}\Gamma_\text{es})\,\rho(u')\frac{dr'}{du'}\,du'\;.
	\end{equation}
	Discretisation is then
	\begin{equation}
		X_i=\sum_i^N\frac{x_i+x_{i+1}}{2}\frac{u_{i+1}-u_i}{u_i^2}\;.
	\end{equation} 
	
\end{appendix}
\end{document}